\documentclass{article}
\usepackage{arxiv}
\usepackage[T1]{fontenc}
\usepackage{microtype}

\usepackage{amsmath}
\usepackage{amssymb}
\usepackage{amsthm}
\usepackage{graphicx}
\usepackage{booktabs}
\usepackage{placeins}
\usepackage{algorithmic}
\usepackage[nocompress]{cite}
\usepackage{hyperref}
\hypersetup{
  pdftitle={Causal Overtaking in Fair Transaction Ordering},
  pdfauthor={Haozhe Feng, Di Wu, and Jian Liu}
}
\newtheorem{theorem}{Theorem}

\newtheorem{definition}[theorem]{Definition}
\newcounter{algorithm}
\newcounter{finding}

\begin{document}

\date{}
\title{Causal Overtaking in Fair Transaction Ordering}

\author{%
  Haozhe Feng, Di Wu\thanks{Corresponding author (\texttt{wu.di@zju.edu.cn}).}, and Jian Liu\\
  Zhejiang University
}

\maketitle
\setcounter{footnote}{0}

\begin{abstract}
Fair transaction ordering protocols aim to mitigate front-running using replica-local arrival orders, but prior work often assumes a single submission origin for each responsive transaction. We identify \emph{causal overtaking}, in which corrupted replicas observe a target and submit identical signed copies of a responsive transaction through multiple client entries. The copies’ first arrivals at different replicas can collectively provide ordering support that no individual path achieves, allowing the response to execute before its target. We formalize a class of receipt-order protocols, including Themis, characterize the timing and network conditions under which the attack is feasible, and show that delayed gossip does not rule it out. We further prove that, under our network model, histories requiring opposite transaction orders can produce identical receipt evidence. Thus, receipt-only mechanisms cannot always preserve earlier-submission priority while preventing responsive overtaking. We propose a mitigation and analyze its assumptions and limitations. Finally, we evaluate the attack and mitigation on Themis through simulations driven by AWS and WonderProxy latency measurements.
\end{abstract}

\section{Introduction}
\label{sec:introduction}

Transaction order affects the prices users pay on decentralized exchanges.
An adversary that observes a pending large purchase can buy the same asset
first, leaving the original buyer with a higher execution price. The
adversary can then sell at a profit after the buyer's trade pushes the
price up further~\cite{daian2020flashboys,torres2021frontrunnerjones}.
Receive-order-fair protocols seek to limit this advantage by constraining
execution according to transaction arrivals observed across
replicas~\cite{kelkar2020aequitas,kursawe2020wendy,kelkar2023themis}.
These arrivals depend on the delivery paths available to each sender,
so a causally later response need not arrive later.

Themis studies this network advantage by fixing a target source and an
observing node, then counting replicas that a follower submitted by that
node can reach first~\cite{kelkar2023themis}. This comparison rules out
overtaking when fixed delays satisfy the triangle inequality. Measured
Internet delays can violate that inequality, making an indirect path
faster than direct delivery~\cite{wang2007tiv}. When several client entries
submit the same follower, their favorable paths can reach different
replicas first.

\begin{figure}[t]
  \centering
  \includegraphics[width=\textwidth]{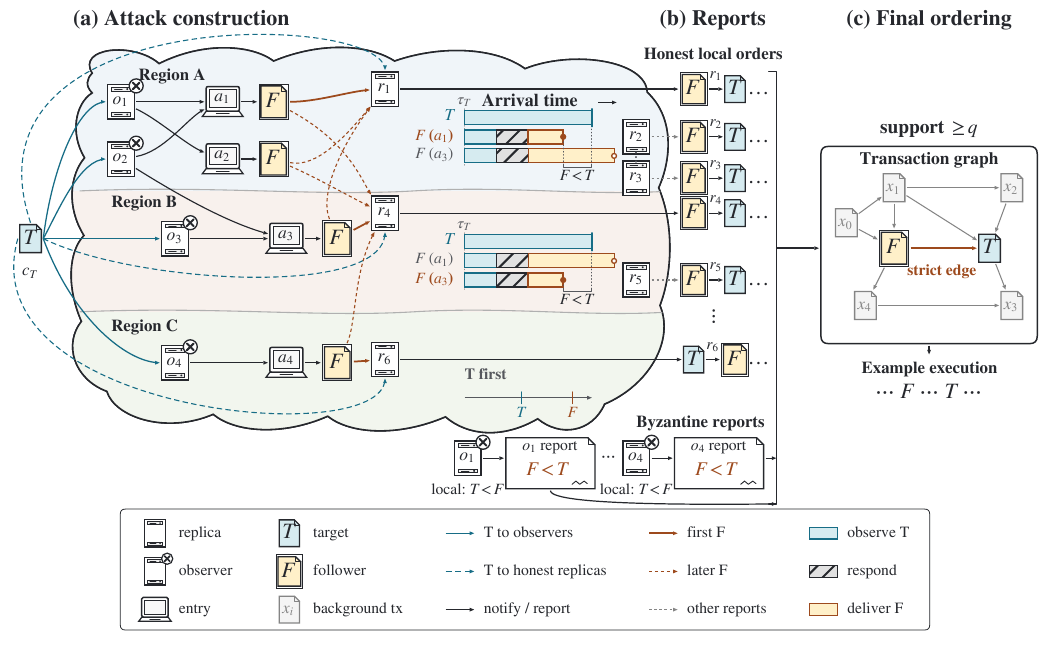}
  \caption{Causal overtaking through combined response paths.
  (a) Observers run the same program to submit identical copies of $F$
  through multiple entries after observing $T$. Different entries reach
  different replicas first. Timing bars start at $T$'s broadcast.
  (b) Honest receipt orders and Byzantine claims contribute ordering evidence.
  (c) Support reaching threshold $q$ creates $F\rightarrow T$, with one
  successful execution shown.}
  \label{fig:multi-entry-overview}
\end{figure}

We define \emph{causal overtaking} as the execution of a responsive follower
before the target that triggered its submission. After observing a pending target
$T$, corrupted replicas trigger ordinary client entries at multiple
locations to submit identical signed copies of a responsive follower $F$.
Because all copies share one transaction identifier, each honest replica
records only the first valid copy it receives, so different replicas can
report copies from different entries. Their reports combine these delivery
advantages into support for ordering $F$ before $T$
(Figure~\ref{fig:multi-entry-overview}).

First-receipt reports reveal when copies arrived but omit how the entries
came to submit them. Under our network model, we construct a responsive
history whose authenticated receipt evidence is identical to that of an
earlier submission of $F$. Because these histories require opposite
execution orders, a receipt-only rule cannot always preserve
earlier-submission priority while preventing responsive overtaking.

We therefore introduce regional ordering, an empirical mitigation that
changes how receipt evidence is produced and aggregated. Replicas
promptly forward transactions within regions with short internal links,
and the committee combines one ordering opinion per region before
finalization. This deployment assumes a bound on corruption within each
region. For evaulation, we replay Themis using AWS, WonderProxy, and ESnet latency
measurements~\cite{park2025ambush,wonderproxy2020day,esnetPerfsonar}.
At $n=101$, the response executes first in 79.1\% of AWS workloads and
89.8\% of WonderProxy workloads. ESnet replays at $n=21$ show overtaking
with measured one-way delays.\footnote{ Code is available at
\url{https://github.com/gdrgrdg12/Causal-overtaking}.}

For mitigation, matched five-region AWS and WonderProxy replays at $n=21$, with one corrupted
observer per region, show that regional ordering roughly halves Themis's
final attack rates for the fixed reference attack while reversing fewer
than 2\% of normal transaction pairs with unanimous honest receipt order.
We then test a finite family of adaptive policies, including strategies
selected on historical data against the frozen regional rule.

\subsection{Our Contributions}
\begin{itemize}
  \item We define causal overtaking and characterize how response delays
        and multiple submission paths determine honest ordering support,
        instantiating the analysis for eight protocol comparisons.
        
  \item We prove that, in our model, receipt-only rules that preserve
      earlier-submission priority cannot prevent all causal overtaking,
      motivating efforts to reduce its frequency.

  \item We propose regional ordering and quantify its effects on finalized overtaking and normal transaction order in trace-driven Themis simulations.
\end{itemize}

\section{Preliminaries}

\subsection{Transaction Ordering}

In programmable blockchains and decentralized finance, the relative order of
otherwise valid transactions can determine arbitrage, slippage, liquidation,
or priority access to shared state~\cite{werner2022sokdefi,
heimbach2022sokreordering}. An adversary that observes a
pending transaction can strategically inject adjacent transactions to extract
value, yielding the front-running and back-running behaviors associated with
maximal extractable value
(MEV)~\cite{daian2020flashboys,torres2021frontrunnerjones}.

Where transactions are ordered depends on the underlying system
architecture. In Layer 1 (L1) blockchains, transaction order is established
through the base chain's own block-construction and consensus
mechanisms~\cite{bano2019sokconsensus}. Layer 2 (L2) systems process
transactions outside the L1 and use the L1 to settle their resulting state
under the L2 protocol's verification and dispute
rules~\cite{gudgeon2020soklayer2}. Rollups, a prominent L2 approach,
aggregate transactions into batches and publish transaction data and state
commitments to L1. A \emph{sequencer} is the component that receives
transaction submissions and determines their execution order within the
rollup~\cite{motepalli2023decentralizedsequencers}.

A single sequencer can use this authority to position its own transactions
around pending submissions. Distributed ordering assigns this task to a
committee of nodes, whose members need a verifiable rule to constrain
their ordering discretion when agreeing on the transaction
sequence~\cite{kelkar2020aequitas,motepalli2023decentralizedsequencers}.

\subsection{Order Fairness}
\label{sec:order-fairness}

Transaction order-fairness constrains the order a protocol may assign to
transactions. Receive-order protocols ground this constraint in local receipt
evidence, seeking to preserve a relative arrival order supported by
sufficiently many replicas~\cite{kelkar2020aequitas}. Let $n$ denote the committee size
and $f$ the maximum number of Byzantine replicas, which may deviate
arbitrarily from the protocol. Each replica $r$ records a local transaction
order $L_r$, with $X\prec_rY$ denoting receipt of $X$ before $Y$. Because
network delays differ, replicas may report different local orders, but
their relative preferences provide evidence without synchronized clocks.

Majorities may prefer $X$ before $Y$, $Y$ before $Z$, and $Z$ before $X$,
forming a \emph{Condorcet cycle}. Any execution sequence must then go
against at least one of these preferences.

To accommodate such cycles, Aequitas's \emph{block-order-fairness}
formulation provides the basis for
\emph{batch-order fairness} in Themis~\cite{kelkar2020aequitas,kelkar2023themis}.
The parameter $\gamma$ specifies how much receipt support triggers the
fairness constraint.
A sufficiently supported arrival order requires the batch containing $X$
to appear no later than the batch containing $Y$. A common batch satisfies
this requirement while leaving the pair's internal order to a separate rule.
These batches are logical groups for the fairness guarantee and may span
multiple blocks or rollup submissions. We write $X\prec_{\mathsf{out}}Y$
when $X$ precedes $Y$ in the eventual execution.

Content-dependent front-running combines early visibility of a pending
transaction with the ability to execute a follower ahead of it. Protections
can restrict either visibility before ordering or the admissible transaction
sequence. \emph{Input causality} requires a follower derived from a pending
transaction's contents to remain behind that transaction. Concealing contents
until the order is fixed provides one way to enforce this
property~\cite{cachin2001securebroadcast,duan2017causalbroadcast}.
Receive-order fairness uses observable arrivals to constrain the sequence
while allowing replicas to inspect pending transactions.

\subsection{Fair Ordering Protocols}
\label{sec:receive-order-interface}
\label{sec:themis-scope}

Aequitas incorporates order-fairness into Byzantine fault-tolerant (BFT)
consensus~\cite{kelkar2020aequitas}. BFT consensus maintains compatible
committed logs at honest replicas and ensures eventual inclusion of valid
transactions under its fault and network assumptions
~\cite{lamport2019byzantine,schneider1990smr}. In leader-based protocols such
as PBFT and HotStuff, replicas validate and vote on a leader's proposed log
extension~\cite{castro1999pbft,yin2019hotstuff}. Fair ordering adds a
verifiable constraint derived from transaction receipts.

Aequitas, Themis, and Quick Order Fairness (QOF) aggregate local orders into
graph constraints between transactions
~\cite{kelkar2020aequitas,kelkar2023themis,cachin2022quickorderfairness}.
QOF bases its fairness guarantee on the difference between opposing
preferences. SpeedyFair separates fair-order computation from consensus
to improve throughput~\cite{mu2024speedyfair}.
These designs group cycles into \emph{strongly connected
components} (SCCs), maximal sets of mutually reachable vertices, and order
the components by their dependencies. Within a component, Themis's
\emph{Unspool} selects a Hamiltonian cycle, which visits every transaction
once, and opens it into a deterministic execution sequence.

Other designs change which transactions are compared or how evidence is
collected. Rashnu and DAG of DAGs (DoD) constrain pairs with conflicting
accesses to shared state~\cite{nagda2024rashnu,nagda2025dod}. DoD and
Herring collect ordering evidence through DAG-based consensus
~\cite{nagda2025dod,putnik2026herring}. FairDAG offers a relative-order
variant (RL) and a variant based on numerical arrival indicators (AB)
~\cite{kang2025fairdag}.

Aggregation can also use information beyond a pair's current preferences.
AUTIG accumulates support across rounds, while Tilikum-BOF preserves
previously established edge directions
~\cite{ren2025autig,segalini2026tilikum}. Ranked Pairs processes preferences
by strength and accepts edges that preserve acyclicity
~\cite{ramseyer2024streamingfairness}. Auncel uses transaction positions in
local orders to compute priorities~\cite{chen2024auncel}, while
Pomp\={e} orders transactions by the median of replica-supplied
timestamps~\cite{zhang2020pompe}.

Ambush exploits temporary differences in transaction visibility to create
cyclic dependencies that let a follower execute before its
target~\cite{park2025ambush}. Its proposed \emph{delayed gossip} mitigation
has replicas probabilistically enqueue transactions for peers and relay
each queue once it contains a prescribed number of transactions.

\section{System and Threat Model}
\label{sec:system-threat}

We specify the network assumptions and adversary capabilities used to
analyze causal overtaking. We then classify fair ordering protocols by
the evidence their comparison rules use and define attack success through the causal relation
between transaction submissions and their final execution order.

\subsection{System and Network Model}
\label{sec:parties}
\label{sec:network-timing}
\label{sec:first-copy-semantics}

We consider fair ordering protocols that use replica-local arrival orders as
ordering evidence. The system comprises a permissioned committee $\mathcal R$
of $n$ replicas and a set of clients. A \emph{replica} records local
transaction arrivals, reports their order, and participates in consensus on
the execution log while maintaining a local \emph{mempool} of valid
transactions awaiting finalization. Replicas may have different mempool
contents and arrival orders.
A \emph{client} is a user or application process that creates, signs, and
submits transactions for execution, without participating in committee
consensus.

Replicas can inspect the pending transactions they have locally received.
Clients know their own transactions, but the committee exposes a
\emph{write-only} submission interface: clients can submit transactions
directly or through relays without gaining read access to replica mempools
or a public feed of other clients' pending transactions. Honest replicas keep
their pending state within the committee's prescribed dissemination process.

Communication operates across two planes. The \emph{transaction plane}
carries client submissions and replica-to-replica gossip. The
\emph{consensus plane} carries authenticated ordering evidence and messages
used to agree on and finalize the execution log.
We assume unforgeable signatures, with clients signing transactions and
replicas signing their reports and consensus messages.

Communication is partially synchronous~\cite{dwork1988partialsynchrony}.
Before the global stabilization time (GST), delays are unbounded. After GST,
honest consensus messages and ordinary transaction transmissions obey their
stated bounds. For a transmission with a specified release time, delivery
path, and recipients, the network schedule determines when it arrives.
Relaying, queueing, and gossip contribute to this delay, while deliveries
must respect causal release, channel order, and the stated bounds.

Global time $t$ describes the execution for analysis. The ordering protocol
consumes replica-local arrival orders without synchronized clocks. For a sender $u$
and recipient replica $r$, let $d_t(u,r)$ denote the effective delay from a
transmission initiated at time $t$ until $r$ records it. This delay includes
propagation, relay, queueing, and local processing. Delays may differ by
direction and vary over time.

Let $\mathsf{arr}_r(X)$ denote the time of transaction $X$'s first valid
receipt at replica $r$, which determines its position in the local arrival
order. If an honest client $c_T$ broadcasts a target transaction $T$ at time
$\tau_T$, then $\mathsf{arr}_r(T)=\tau_T+d_{\tau_T}(c_T,r)$.
Simultaneous arrivals are resolved deterministically.

\subsection{Adversary Capabilities}
\label{sec:client-capability}

We consider a static adversary controlling a set
$\mathcal B\subseteq\mathcal R$ of at most $f$ corrupted replicas.
The remaining replicas $\mathcal H=\mathcal R\setminus\mathcal B$ are honest.

Corrupted replicas can inspect their local mempools, disclose received
transactions, and misstate their own signed reports.

An \textbf{observer} is a corrupted replica whose local receipts provide
pending-transaction observations to the adversary. The \emph{observer set}
is $\mathcal O\subseteq\mathcal B$. An observer learns $T$ through its
replica's first valid local receipt and may disclose the transaction contents,
submission metadata, or receipt event. The protocol may accept reports from replicas other than those used for
observation.

A \textbf{client entry} is a client-side submission endpoint that sends
transactions to a fixed, nonempty set of accessible replicas, directly or
through a relay. The adversary controls a finite set $\mathcal A$ of entries
whose number is independent of the replica corruption bound $f$. Entries
learn about pending targets only through observer notifications. An observer
and an entry need not be co-located.

An entry's location is its \emph{submission origin}. A \emph{response path}
includes delivery of the target to an observer, notification of an entry,
and delivery of that entry's follower copy to a replica.

The adversary can predeploy response programs and provision shared
transaction parameters and signing keys at its controlled locations.

An observer may notify multiple entries after receiving $T$. The adversary
chooses which available observers and entries participate and when each
causally enabled entry submits. We hold network scheduling outside this
adversary's control to isolate the effect of submission and report choices.
This endpoint adversary is weaker than the delivery scheduler allowed in
standard partial-synchrony security models. Its decisions use observations
available at submission time.

\noindent\textbf{Observer--entry pipeline.}
We write $F$ for the adversarial follower transaction. After recording $T$,
observer $o$ spends $\sigma_o\geq0$ on detection, selection, construction,
or signing before notifying a linked entry $a$. A notification initiated at
time $t$ incurs an effective delay $\kappa_t(o,a)$, including any transfer of
data needed to submit $F$, under the same partial-synchrony assumptions as
transaction transmissions. Entry $a$ then spends $\rho_a\geq0$ processing
the notification before invoking its submission interface. This permits
constructing $F$ after observation or releasing a previously prepared signed
$F$. All participating entries submit byte-identical copies under the same
client signing identity. When several observers can notify one entry, its
earliest notification, followed by local processing, determines when it can
first release $F$.
\begin{table}[tb]
  \centering
  \caption{System and adversary notation.}
  \label{tab:model-parties}
  \small
  \begin{tabular}{@{}l p{0.64\columnwidth}@{}}
    \toprule
    Symbol & Meaning \\
    \midrule
    $n,f$ & Committee size and replica corruption bound \\
    $\mathcal R$ & All committee replicas \\
    $\mathcal B,\mathcal H$ & Corrupted and honest replicas \\
    $\mathcal O\subseteq\mathcal B$ & Corrupted replicas acting as observers \\
    $\mathcal A$ & Adversarial client submission entries \\
    $\mathcal S\subseteq\mathcal R$ & Reporters for one ordering decision \\
    $b$ & Number of corrupted reporters in $\mathcal S$ \\
    $T,F$ & Honest target and adversarial follower \\
    $c_T,\tau_T$ & Target client and its broadcast time \\
    $\mathsf{arr}_r(X)$ & First valid receipt time of $X$ at replica $r$ \\
    $d_t(u,r)$ & Effective transmission delay from $u$ to replica $r$ \\
    $\kappa_t(o,a)$ & Notification delay from observer $o$ to entry $a$ \\
    $\sigma_o,\rho_a$ & Observer and entry processing delays \\
    \bottomrule
  \end{tabular}
\end{table}

Table~\ref{tab:model-parties} summarizes the main model notation.

\subsection{Protocol Classification}
\label{sec:ordering-model}

\begin{table}[t]
  \centering
  \caption{Properties of protocol comparison rules.
  $\checkmark$: satisfied. $\times$: additional evidence required.}
  \label{tab:protocol-coverage}
  \small
  \setlength{\tabcolsep}{4pt}
  \renewcommand{\arraystretch}{1.12}
  \begin{tabular}{@{}p{0.22\textwidth}ccc p{0.42\textwidth}@{}}
    \toprule
    Protocol & Ordinality & \shortstack{Pair\\locality}
      & \shortstack{History\\independence} & Comparison / additional evidence \\
    \midrule
    Themis / SpeedyFair~\cite{kelkar2023themis,mu2024speedyfair}
      & $\checkmark$ & $\checkmark$ & $\checkmark$
      & Complete initial comparison from accepted local orders. \\
    Rashnu / DoD~\cite{nagda2024rashnu,nagda2025dod}
      & $\checkmark$ & $\checkmark$ & $\checkmark$
      & Initial conflicting pair. DoD uses a common worker, a process handling a transaction subset. \\
    Herring~\cite{putnik2026herring}
      & $\checkmark$ & $\checkmark$ & $\checkmark$
      & Complete initial comparison with both transactions retained in one graph. \\
    FairDAG-RL~\cite{kang2025fairdag}
      & $\checkmark$ & $\checkmark$ & $\checkmark$
      & Complete committed comparisons at the pair's first orientation. \\
    Aequitas~\cite{kelkar2020aequitas}
      & $\checkmark$ & $\checkmark$ & $\checkmark$
      & Complete agreed first-in-first-out (FIFO) logs. Native asynchronous $\gamma=1$ requires unanimous preference. \\
    QOF~\cite{cachin2022quickorderfairness}
      & $\checkmark$ & $\checkmark$ & $\checkmark$
      & Complete FIFO prefixes, including both directional edge tests. \\
    \midrule
    AUTIG~\cite{ren2025autig}
      & $\checkmark$ & $\checkmark$ & $\times$
      & Edge decisions incorporate support accumulated across rounds. \\
    Tilikum-BOF~\cite{segalini2026tilikum}
      & $\checkmark$ & $\checkmark$ & $\times$
      & Edge decisions retain previously established directions. \\
    Ranked Pairs~\cite{ramseyer2024streamingfairness}
      & $\checkmark$ & $\times$ & $\checkmark$
      & Complete-input aggregation checks paths through other transactions. \\
    Auncel~\cite{chen2024auncel}
      & $\checkmark$ & $\times$ & $\checkmark$
      & Complete-input ranking uses positions in the full local orders. \\
    Pomp\={e}~\cite{zhang2020pompe}
      & $\times$ & $\times$ & $\checkmark$
      & Median comparison with fixed timestamp response sets. \\
    FairDAG-AB~\cite{kang2025fairdag}
      & $\times$ & $\times$ & $\checkmark$
      & Assigned-indicator comparison with complete committed values. \\
    \bottomrule
  \end{tabular}
\end{table}

To analyze how earlier arrivals affect a protocol, we first determine
which evidence its comparison rule uses. The three properties below
identify when a pair's current receipt preferences suffice and when the
analysis needs transaction history, background orders, or numerical
timestamps as well.
Fix the protocol parameters, the identities supplying
accepted evidence, and an eligible transaction pair $(X,Y)$. For graph-based
rules, the decision is the direct edge orientation. For ranking rules, it
is the relative priority assigned to the pair. Admission and delivery
conditions are specified separately.

\noindent\textbf{Ordinality.}
A comparison is ordinal if, for fixed retained comparison state, changing
receipt-time values while preserving every replica's complete local
transaction order leaves the decision unchanged. Arrival inequalities then
suffice to describe its timing input.

\noindent\textbf{Pair locality.}
A comparison is pair-local if the current binary preferences between $X$
and $Y$, together with any retained history of that pair, determine its
decision. Holding that pair history and those preferences fixed, changes to
receipt-time values or background transaction positions leave the direct
relation unchanged. This permits analyzing the pair through its local
supporters.

\noindent\textbf{History independence.}
The current accepted evidence determines the decision regardless of
earlier accumulated support or retained orientations. A single collection
of reports then supplies the comparison input.

The first six rows of Table~\ref{tab:protocol-coverage} cover eight
constructions whose stated comparisons satisfy all three properties.
Their current pairwise preferences supply the evidence needed for a direct
edge decision. Signed reports, agreed FIFO log prefixes, and committed DAG
blocks authenticate and select these preferences. Each prefix contains a
sender's initial broadcasts in order. The transaction graph built from
this evidence may contain cycles.
Data-dependent protocols additionally select pairs with conflicting
accesses to shared state.

If a comparison lacks one of these properties, we must supply the
corresponding evidence. For AUTIG and Tilikum-BOF, this means a realizable
history of support or orientations.
Ranked Pairs and Auncel require background transaction orders, which affect
cycle checks and position weights. Timestamp and absolute-indicator rules
require the actual numerical values. An attack against these rules must
produce those inputs through permitted submissions and show how the
resulting comparison affects execution.

For the eight comparisons satisfying all three properties, we express the
accepted preferences through a common support count. Let the \emph{reporter
set} $\mathcal S\subseteq\mathcal R$ contain the identities whose local
orders supply the evidence, and write $m=|\mathcal S|$. We use complete
comparisons: each selected report contains both transactions in a strict
order and contributes once. The protocol determines evidence admission and
the point at which the pair is resolved. Write
$b=|\mathcal B\cap\mathcal S|$ for the selected corrupted reporters.
Corrupted observers may lie outside an entirely honest reporter set.
The accepted support for $X$ before $Y$ is
\[
  w_{\mathcal S}(X,Y)
    =|\{r\in\mathcal S:X\prec_rY\}|.
\]
A \emph{direct edge} is formed by the comparison rule before SCC
finalization. It is \emph{strict} when its direction is selected
independently of tie breaking and the opposite edge is excluded.
In the eight stated
comparisons, reporters contribute equally and increasing forward support
preserves a strict forward decision. These rules therefore have an integer
threshold $q$ for which a strict edge forms exactly when
$w_{\mathcal S}(X,Y)\geq q$. For a support floor $h$ combined with a strict
majority, the threshold is
\[
  q=\max\{\lceil h\rceil,\lfloor m/2\rfloor+1\}.
\]
Appendix~\ref{app:protocol-thresholds} derives the native thresholds,
specifies the required report and pair conditions, and gives a support
pattern meeting all eight thresholds.
For graph-based finalization that respects dependencies across SCCs, a
preserved path from $X$ to $Y$ with no path back places $X$ before $Y$
when both are delivered. Within an SCC, the protocol's internal ordering
rule determines the execution outcome.

\subsection{Fairness and Attack Objective}
\label{sec:causal-goal}

Receive-order fairness constrains execution order using replica receipts.
We use an honest-replica parameterization of batch-order fairness,
adapted from Themis~\cite{kelkar2023themis}, with fairness parameter
$1/2<\gamma\leq1$. If every honest replica
receives $X,Y$ and at least $\gamma(n-f)$ honest replicas receive $X$
first, the batch containing $X$ must appear no later than the batch
containing $Y$. Appendix~\ref{app:protocol-thresholds} converts this
parameter to Themis's original all-replica definition.

The adversary aims to place a responsive follower $F$ ahead of an honestly
submitted target $T$ in the finalized execution. We consider valid
target--follower pairs whose relative order affects validity, output, or
economic effect. Execution precedence denotes log order in serial execution or enforced
precedence of the pair in parallel execution.

\noindent\textbf{Responsive constraint.}
Let $\mathsf{submit}(X)$ denote the first protocol-facing submission
invocation of transaction $X$, and let $\prec_{\mathsf{hb}}$ denote causal
\emph{happens-before} precedence. Write $\mathsf{observe}_o(T)$ for observer
$o$'s first valid receipt of $T$ and $\mathsf{submit}_{o,a}(F)$ for the
submission invocation at entry $a$ triggered by that receipt. An execution
is \emph{responsive} only if no copy of $F$ is submitted before a triggering
observation and, for every entry $a$ originating a copy, there is an observer
$o\in\mathcal O$ able to notify $a$ such that
\[
  \mathsf{submit}(T)\prec_{\mathsf{hb}}\mathsf{observe}_o(T)
  \prec_{\mathsf{hb}}\mathsf{submit}_{o,a}(F).
\]
The observation must trigger the invocation: holding the adversarial state,
coins, and execution prefix fixed, the entry would not submit at that point
without the observation.

\begin{definition}[Causal overtaking]
\label{def:causal-overtaking}
Consider a valid target $T$ submitted by an honest client and a valid
follower $F$ satisfying the responsive constraint. A causal overtaking
attack succeeds in a legal execution when both transactions are finalized
for execution and $F\prec_{\mathsf{out}}T$.
\end{definition}

The responsive constraint covers both a follower constructed from $T$'s
contents and a prepared transaction released upon observing $T$.
In either case, honest replicas can receive and truthfully report $F$
first, and their receipts can satisfy the protocol's receive-order fairness
requirement even when the responsive follower executes first.
We measure strict-edge formation separately to identify executions where
the aggregated evidence itself favors the responsive follower.

\section{Causal Overtaking in Receive-Order Protocols}
\label{sec:multi-entry-attack}

Under the model in Section~\ref{sec:system-threat}, an attacker must
deliver the follower early enough to affect honest reports, then turn
those reports into an execution advantage. We describe the attacker's submission
procedure, derive the available timing margins and support, and follow
the resulting comparison through protocol finalization.

\subsection{Attack Construction}
\label{sec:attack-construction}

The adversary uses corrupted replicas to observe $T$ and ordinary client
entries to submit the same signed $F$.\footnote{Publicly readable mempools or transaction
feeds let client entries observe $T$ directly and submit $F$ without
corrupting replicas. Appendix~\ref{app:public-observation} analyzes this
variant.} Before observing $T$, it deploys the same deterministic response
program at its observer locations and configures the entries to relay the
program's output.
The program constructs $F$ from $T$ using shared transaction parameters,
including the nonce and fees. A common adversarial client key, deterministic
signing, and identical serialization let each observer independently
produce the same signed bytes~\cite{pornin2013deterministicecdsa}.
Upon receiving $T$ locally, an observer runs the program and notifies one
or more entries with the resulting $F$. For a prepared follower, the
programs store the same signed transaction and release it upon observation.
Construction and signing contribute to the observer's processing delay
$\sigma_o$.

An entry submits through the ordinary client interface after processing
its first valid notification, using the earliest available observation
when several observers can notify it.
Each observer can also activate several entries, so different replicas
may receive their first copies of $F$ from different entries. Honest
replicas report these arrivals accurately, while corrupted replicas may
contribute additional admissible $F$-first reports.

Figure~\ref{fig:multi-entry-overview} illustrates the construction and its
ordering evidence. Panel~(a) shows different entries supplying the earliest
copies of $F$ to different honest replicas, with timing bars comparing their
complete response paths. Panel~(b) collects truthful local orders alongside
possible Byzantine reports. In panel~(c), sufficient accepted support
creates $F\rightarrow T$, with one final execution placing $F$ before $T$.

\subsection{Attack Analysis}
\label{sec:attack-capacity}

For an honest replica to receive $F$ strictly before $T$, the complete
response path must fit within the target's delivery time. Fix an execution in which $c_T$ broadcasts
$T$ at $\tau_T$, with response path
$c_T\rightarrow o\rightarrow a\rightarrow r$. Entry $a$ can submit only
after observer $o$ receives $T$, prepares $F$, and notifies it. Including
the entry's local processing, its earliest release offset relative to
$\tau_T$ is
\[
  \begin{aligned}
  \alpha_a=\min_o\bigl\{&
    \mathsf{arr}_o(T)-\tau_T+\sigma_o\\[-0.4ex]
    &+\kappa_{\mathsf{arr}_o(T)+\sigma_o}(o,a)+\rho_a\bigr\},
  \end{aligned}
\]
where the minimum ranges over observers able to notify $a$, with
$\min\varnothing=+\infty$. In the construction, every entry with finite
$\alpha_a$ submits at $\tau_T+\alpha_a$. Let $\mathcal A_r$ contain these
entries whose dissemination reaches honest replica $r$. Its first receipt is
\[
  \mathsf{arr}_r(F)=\tau_T+
  \min_{a\in\mathcal A_r}\{
       \alpha_a+d_{\tau_T+\alpha_a}(a,r)\}.
\]
The minimizing entry may differ across replicas because each replica
keeps its own earliest copy.

If transmission delays stay fixed as release times vary, we can express
the entry's permitted timing as an interval. Write $d(c_T,r)$ and $d(a,r)$
for those delays, and let $s_a$ be an entry's chosen release offset. A
causally enabled submission reaches $r$ before $T$ exactly when
\[
  s_a\in[\alpha_a,\ d(c_T,r)-d(a,r)).
\]
The lower endpoint is the earliest causally enabled release. At the upper
endpoint, the copy would arrive together with $T$, so overtaking requires
an earlier release. If the interval is empty, the entry cannot respond
quickly enough along this path.

For time-varying delays, the corresponding condition is
$s_a\geq\alpha_a$ and
$s_a+d_{\tau_T+s_a}(a,r)<d_{\tau_T}(c_T,r)$.
We use realized delays to evaluate the construction at $s_a=\alpha_a$.
For distinct first-receipt times, define the entry's \emph{winner set}
\[
  \begin{aligned}
  \mathsf{Win}(a)=\{r\in\mathcal H:\;&a\in\mathcal A_r,\\[-0.4ex]
    &\alpha_a+d_{\tau_T+\alpha_a}(a,r)<d_{\tau_T}(c_T,r)\}.
  \end{aligned}
\]
The union of these sets gives the honest replicas whose receipts favor $F$.
Its cardinality is the attack's \emph{capacity} in this execution:
\[
  C=\left|\bigcup_{a\in\mathcal A}\mathsf{Win}(a)\right|.
\]
Each replica contributes once, so an additional entry increases capacity
only when it reaches honest replicas outside the existing union before $T$.

\label{sec:attack-aggregation}
The protocol accepts evidence from the reporter set $\mathcal S$ defined
in Section~\ref{sec:ordering-model}. The realized honest support is
\[
  v=\left|\mathcal S\cap
       \bigcup_{a\in\mathcal A}\mathsf{Win}(a)\right|.
\]
For a complete accepted comparison of an eligible pair $(F,T)$, suppose the
$b$ selected corrupted reporters supply admissible $F$-first evidence.
Truthful honest reports then give $w_{\mathcal S}(F,T)=v+b$, so the native
comparison rule creates a strict $F\rightarrow T$ edge exactly when
\[
  v+b\geq q.
\]
When the completed graph preserves this edge between distinct SCCs,
topological execution places $F$'s component before $T$'s. Delivery of both
transactions in this order constitutes a successful causal overtaking attack.

We next characterize the support attainable by choosing entries and release
times. Fix the available observer notifications and submission paths, with
processing and transmission delays independent of release choices and the
active entry set. The target's arrivals and the release offsets $\alpha_a$
therefore remain fixed. For each honest replica $r$, define the best arrival
margin over the available entries as
\[
  M_r=\max_{a\in\mathcal A_r}
       \{d(c_T,r)-\alpha_a-d(a,r)\}.
\]
Take $M_r=-\infty$ when no available entry reaches $r$. Let $\delta\geq0$
be a common additional processing delay at the entries. A submission policy
may activate any subset of entries and choose a release offset
$s_a\geq\alpha_a+\delta$ for each active entry.

\begin{theorem}[Attack feasibility]
\label{thm:attack-feasibility}
Fix an eligible complete comparison, its accepted reporters $\mathcal S$,
threshold $q$, and $b$ corrupted contributions independently of entry
activation and release choices. Under the fixed delays above, assume the
earliest permitted releases produce no target--follower receipt ties.
The maximum honest support over all permitted submission policies is
\[
  v_{\max}(\delta)
    =|\{r\in\mathcal S\cap\mathcal H:M_r>\delta\}|.
\]
Activating all available entries and submitting at
$s_a=\alpha_a+\delta$ attains this maximum. Let $k=q-b$.
For $1\leq k\leq|\mathcal S\cap\mathcal H|$, write $M_{(k)}$ for the
$k$th largest margin among the accepted honest reporters. A permitted
submission policy creating a strict $F\rightarrow T$ edge exists exactly when
\[
  0\leq\delta<M_{(k)}.
\]
\end{theorem}

\begin{proof}
For any permitted policy, a copy submitted through entry $a$ reaches $r$
no earlier than $\tau_T+\alpha_a+\delta+d(a,r)$. Receipt before $T$
therefore requires $\delta<M_r$. Every honest supporter belongs to the
set counted by $v_{\max}(\delta)$, establishing the upper bound.

To attain the bound, each available observer notifies its linked entries
after processing $T$. Each entry submits upon its first valid notification
after completing its local processing and the additional delay $\delta$.
These actions use received notifications and local timers, so every
submission satisfies the responsive constraint. The resulting first receipt
at $r$ is
\[
  \mathsf{arr}_r(F)=\tau_T+\delta+
    \min_{a\in\mathcal A_r}\{\alpha_a+d(a,r)\}.
\]
It precedes $\mathsf{arr}_r(T)$ exactly when $\delta<M_r$. Thus one
submission policy attains the upper bound at all reporters simultaneously.
It supplies at least $k$ honest preferences exactly when
$\delta<M_{(k)}$. Adding the $b$ corrupted preferences reaches $q$ in
that interval. Outside it, the maximum honest support is below $k$.
\end{proof}

At $\delta=0$, the construction attains $v=v_{\max}(0)$. Applying the
same argument to all honest replicas makes $C$ the maximum honest receipt
capacity under these fixed delays. A positive $M_{(k)}$ gives the additional
processing time available while retaining enough support for an edge.
If $k\leq0$, the corrupted contributions already reach the threshold.
If $k>|\mathcal S\cap\mathcal H|$, the total available support is below $q$.

Because a policy may leave entries inactive, adding available entries
cannot reduce the maximum support under these fixed-delay assumptions.
Any gain depends on their notification delays and whether they reach
replicas outside the existing winner sets.

For a matched single-entry comparison, retain $\mathcal S$, the corrupted
reports, and each entry's release and delivery delays. Let
$v_a=|\mathcal S\cap\mathsf{Win}(a)|$. Then $\max_a v_a$ is the support
of the best individual entry, while $v$ counts the combined support.
The combined support reaches the threshold when $v\geq q-b$.
The stronger case in which even the best individual entry falls short is
\[
  \max_{a\in\mathcal A}v_a<q-b\leq v.
\]
This comparison evaluates every available entry separately on the same
execution, including all of its observer notifications.

\subsection{From Evidence to Execution}
\label{sec:attack-applicability}

For the comparisons in Table~\ref{tab:protocol-coverage}, the protocol must
admit both transactions and accept the supporting reports before the edge
can affect their execution. We therefore examine each protocol's admission
rules before following a direct $F\rightarrow T$ edge through graph
finalization.
For Rashnu and DoD, the attacked pair must have conflicting accesses, with
DoD assigning both transactions to corresponding workers across replicas.
Herring selects
an anchor transaction that bounds the graph portion retained for delivery,
and both transactions must survive this selection and filtering.
FairDAG-RL must incorporate their comparisons into committed evidence at
first orientation. Aequitas and QOF wait for enough evidence to resolve
dependencies on transactions beyond the current graph before delivering
its components~\cite{kelkar2020aequitas,cachin2024qofimplementation}.

For rules outside this class, the attacker must also produce the history
or background orders used by the comparison. In Tilikum-BOF, the follower's support must
establish its direction before the reverse edge is inserted. In AUTIG,
current support is added to the pair's earlier weights. In Ranked Pairs,
a candidate $F\rightarrow T$ edge is accepted when the stronger or
previously processed edges leave no path from $T$ back to $F$
\cite{ren2025autig,segalini2026tilikum,ramseyer2024streamingfairness}.

Themis's network-level analysis gives a nonpositive arrival margin in a
metric network when response overhead is zero and the entry shares its
observer's delivery paths~\cite{kelkar2023themis}. Overtaking requires a
positive margin on at least one response path, and different entries can
provide it at different replicas. Ambush's delayed gossip
changes these margins through relaying and queueing~\cite{park2025ambush}.
While target copies wait for a queue's release threshold, direct client
submissions can still overtake them.

Once a preserved $F\rightarrow T$ edge connects distinct SCCs, component
order puts $F$ before $T$ when both are delivered. Background transactions
can instead connect the pair through a return path and place them in one
SCC. Aequitas and QOF can deliver that component as a batch, leaving its
internal execution order to the consuming system
\cite{kelkar2020aequitas,cachin2022quickorderfairness}.
Themis resolves the component into a linear sequence through Unspool.

\phantomsection
\label{sec:within-scc-outcomes}
For a Themis SCC containing both transactions, fix the directed Hamiltonian
cycle selected by Unspool and write its vertex sequence, starting at $F$, as
\[
  (F,X_1,\ldots,X_j,T,Y_1,\ldots,Y_{\ell-j-2}).
\]
Here $\ell$ is the component size and $j$ counts the transactions between
$F$ and $T$. The $Y$ transactions lie on the remaining arc. The selected
cycle may connect $F$ to $T$ through these intermediate transactions.
Opening it after $T$ or any $Y$ yields $F\prec_{\mathsf{out}}T$,
while opening it after $F$ or any $X$ yields the opposite order.

The cut is deterministic. For the proposal's last SCC, the sequence must
end with a \emph{solid} transaction, one present in at least $n-2f$
reports~\cite[Figure~3]{kelkar2023themis}. Subject to this endpoint
constraint, Themis gives cutting a smallest-weight cycle edge as one
possible rule. An edge's weight counts reports supporting its direction.
The edges on each arc determine which eligible cuts yield each
relative order. Background traffic can change component membership,
edge weights, the selected cycle, and which component must end with a
solid transaction. The eligible cuts can therefore change even when
the pair's direct edge stays the same.

\FloatBarrier

\section{Mitigation}
\label{sec:carryover-guard}

We first show why receipt evidence alone cannot always distinguish an
earlier submission from a responsive follower. We then introduce regional
ordering as an empirical mitigation with deployment assumptions and
discuss its ordering tradeoffs.

\subsection{Indistinguishable Submission Histories}
\label{sec:receive-order-limits}
\label{subsec:collision-example}

Consider valid transactions $F,T$ with both execution orders permitted
by the application. Earlier-submission priority requires $F$ before $T$
when $F$ is submitted first. Preventing responsive overtaking requires
$T$ before $F$ when $F$ satisfies the responsive constraint in
Section~\ref{sec:causal-goal}. We call a rule receipt-only when its
decision uses accepted receipt reports and fixed prior ordering state.

We first fix an earlier-submission history, then construct responsive
paths that reproduce its receipts. Consider $N=|\mathcal S|$ honest
reporters and $K$ ordinary client entries at distinct locations, triggered by one corrupted
observer outside $\mathcal S$. Use direct dissemination without
replica gossip for this construction. Let $t_0$ denote the target's submission
time and $\Delta_{\mathsf{tx}}$ bound transaction delivery delays after
GST. Choose $\lambda,\epsilon>0$ with
$\lambda+3\epsilon<\Delta_{\mathsf{tx}}$ and
$t_0-\epsilon>\mathsf{GST}$. For each reporter $r$, fix
$\epsilon_r\in[\epsilon,3\epsilon/2]$. In the earlier history, a client
submits $F$ at $t_0-\epsilon$ with delivery delay
$\lambda+\epsilon+\epsilon_r$ to $r$. The target's delivery delay is
$\lambda+\epsilon_r+\epsilon$. Thus the fixed receipts are
\[
  \mathsf{arr}_r(F)=t_0+\lambda+\epsilon_r,\qquad
  \mathsf{arr}_r(T)=\mathsf{arr}_r(F)+\epsilon.
\]

In the responsive history, the observer receives $T$ at
$t_0+\lambda/2$ and immediately notifies every entry over links of delay
$\lambda/2$. Each entry is therefore notified at $t_0+\lambda$.
Set the direct target-source-to-entry delay to $d(c_T,a)=\lambda$ as well.
Entry $a$ has a fixed processing delay $\rho_a\in[0,\epsilon/2]$ and
releases the same prepared signed $F$ at $t_0+\lambda+\rho_a$.
Construct the entry-to-reporter delays as
\[
  d(a,r)=
  \begin{cases}
    \epsilon_r-\rho_a, & \text{short path},\\
    3\epsilon, & \text{long path}.
  \end{cases}
\]
A short path delivers $F$ at exactly $t_0+\lambda+\epsilon_r$, matching
the earlier history. A long path delivers it after $T$. The term
$-\rho_a$ offsets the entry's processing delay, so a single release time at
each entry matches the prescribed receipts at every reporter reached by
its short paths. All link delays and processing offsets are fixed before
execution and remain the same in both histories. Write
$U=\mathcal S\cap\bigcup_a\mathsf{Win}(a)$ for the reporters covered by
at least one short path, using Section~\ref{sec:attack-capacity}.

\begin{theorem}
\label{thm:indistinguishable-histories}
The responsive history reproduces the fixed earlier history's receipt
orders and timestamps at all selected reporters if and only if
$U=\mathcal S$. Under this condition, a receipt-only rule starting from
the same prior state gives the same output distribution to histories
with opposite ordering requirements.
\end{theorem}

\begin{proof}
If $U=\mathcal S$, every reporter receives a short-path copy at its
prescribed time, and all long-path copies arrive later. If a reporter
is uncovered, its first copy arrives after $T$, differing from the fixed
earlier receipts. All transmission delays are positive and obey
$\Delta_{\mathsf{tx}}$. Keep target delivery, background transactions,
local clocks, signing randomness, prior ordering state, and report delivery
identical. When $U=\mathcal S$, the ordering layer receives the same
authenticated evidence and returns the same output distribution.
Preserving earlier-submission priority requires $F$ first in one history,
while preventing responsive overtaking requires $T$ first in the other.
Neither output satisfies both, and indefinite waiting violates eventual
decision.
\end{proof}

The constructed short paths violate the source--entry--reporter triangle
inequality because
\[
  \begin{aligned}
    d(c_T,a)+d(a,r)&=\lambda+\epsilon_r-\rho_a\\
      &<\lambda+\epsilon_r+\epsilon=d(c_T,r).
  \end{aligned}
\]
For a long path, the two-hop delay is $\lambda+3\epsilon>d(c_T,r)$,
so the triangle inequality holds. In this constructed sparse-path family,
each entry-to-reporter link is
short with probability $0<\pi<1$, independently across entries and
reporters. Here $\pi$ describes these triples for the fixed target.
The probability of matching the fixed earlier receipts is exactly
\[
  \Pr[\text{receipt collision}]
    =\Pr[U=\mathcal S]=\bigl[1-(1-\pi)^K\bigr]^N.
\]
A single entry must supply all $N$ short paths, giving probability $\pi^N$.
With multiple entries, the matching copy can come from a different entry
at each reporter. For fixed $K,N$, the ratio to the single-entry
probability tends to $K^N$ as $\pi\to0$. A union bound gives
$\Pr[U\ne\mathcal S]\leq N(1-\pi)^K$, so the collision occurs with
probability at least $1-\eta$, for $0<\eta<1$, whenever
\[
  K\geq\frac{\ln(N/\eta)}{-\ln(1-\pi)}.
\]
For small $\pi$, the bound requires about $\ln(N/\eta)/\pi$ entries,
all using ordinary client submission interfaces and the same corrupted
observer.

Ordinal rules need only the order of receipts, so short-path delays may
vary over $0<d(a,r)<\epsilon_r+\epsilon-\rho_a$. If background receipts
stay outside the possible arrival window, these changes preserve each
reporter's full local order.

\phantomsection
\label{subsec:collision-implications}
We therefore explore regional ordering, which combines short internal
dissemination paths with one opinion per region to reduce the influence
of complementary delivery advantages across individual replicas.

\subsection{Regional Ordering}
\label{subsec:regional-ordering}

A uniformly random execution order prevents overtaking with probability
50\%, at the cost of reversing the same fraction of normal transaction
pairs. Regional ordering seeks a better tradeoff by grouping replicas
into regions with short internal links and giving each region one ordering
opinion. Co-location reduces external delivery differences, while prompt
forwarding spreads a received transaction within its region. The committee
combines these regional opinions before graph
finalization (Algorithm~\ref{alg:regional-ordering}).

\begin{figure}[!htbp]
  \centering
  \begin{minipage}{\columnwidth}
    \refstepcounter{algorithm}\label{alg:regional-ordering}
    \hrule
    \smallskip
    \noindent\textbf{Algorithm~\thealgorithm}\quad Regional ordering
    \smallskip\hrule\smallskip
    \footnotesize
    \begin{algorithmic}[1]
      \REQUIRE Fixed $\Theta$, base graph $G$, signed receipt evidence, public priorities
      \FOR{each transaction pair $(X,Y)$}
        \FOR{each region in $\Theta$}
          \IF{exactly one direction meets the stable-witness threshold}
            \STATE Set this region's vote to that direction
          \ELSE
            \STATE Take available reports up to the limit by witness priority
            \IF{the sample is empty}
              \STATE This region abstains
            \ELSE
              \STATE Vote by sample majority, using transaction priority for ties
            \ENDIF
          \ENDIF
        \ENDFOR
        \STATE Count the votes cast for each direction
        \IF{exactly one direction $e$ meets the regional threshold}
          \STATE $G\gets(G\setminus\{X\to Y,Y\to X\})\cup\{e\}$
        \ENDIF
      \ENDFOR
      \RETURN Original SCC condensation and Unspool on $G$
    \end{algorithmic}
    \smallskip\hrule
  \end{minipage}
\end{figure}

A deployment fixes the region partition, stable-witness thresholds,
sample size, and regional voting threshold in a configuration $\Theta$
before execution.
The witness thresholds exceed an assumed corruption bound within each
region, a stronger assumption than the global Byzantine budget.

On receiving a complete transaction, an honest replica forwards it once
to its region. Replica $i$ records the first direct receipt
$t_i^{\mathsf{dir}}(X)$ and the first receipt through a regional peer
$t_i^{\mathsf{reg}}(X)$ on the same local clock, filling each record
when that copy type first arrives even if another type has already arrived.
Replicas sign these records for collection with the base reports.

For each transaction pair, a replica is a \emph{stable witness} for $X$
before $Y$ when both receipt types for $X$ precede both receipt types
for $Y$ on its local clock:
\[
  \max\{t_i^{\mathsf{dir}}(X),t_i^{\mathsf{reg}}(X)\}
  <\min\{t_i^{\mathsf{dir}}(Y),t_i^{\mathsf{reg}}(Y)\}.
\]
A missing receipt prevents the replica from establishing stable precedence.
A region votes for the unique direction that meets its stable-witness
threshold. If there is no such direction, the region takes up to the
configured number of available local-order reports in public witness
priority order. It abstains if the sample is empty. For a nonempty
sample, it votes for the majority direction and uses the region's public
transaction priority to break a tie. Each region reuses its priorities
across transaction pairs.

The leader fixes the evidence snapshot and constructs the base transaction
graph using the protocol's original quorum and thresholds, then applies
the regional rule to every transaction pair. An abstaining region
contributes no vote, and the regional threshold stays fixed. When exactly
one direction meets the threshold, the leader replaces the pair's edge
with that direction. Otherwise, it keeps the base edge. Given the same
snapshot of signed regional records and the same public priorities,
replicas can recompute the regional decisions before the original SCC
condensation and Unspool.

\FloatBarrier
\subsection{Discussion and Limitations}
\label{subsec:mitigation-discussion}

Even with a regional $T\rightarrow F$ edge, background transactions may
form a return path from $F$ to $T$, placing both in one SCC where Unspool
can execute $F$ first. One way to address this
is to carry regional support into cycle resolution and give stronger
ordering evidence greater weight when selecting edges to retain.

Regional voting trades some arrival-order preferences for resistance to
responsive overtaking. Edges left unchanged retain the base protocol's
receipt constraints in the graph, while replaced edges use regional
evidence. Normal-order loss measures the cost of this tradeoff on ordinary
traffic. A general fairness guarantee for the combined rule remains to
be formalized.

The mitigation assumes a bound on corruption within each region.
The global fault budget alone permits corruption to concentrate within a region.
With $f=5$, an adversary can control all four members of a four-node region
and determine that region's vote. Deployment must therefore enforce the
regional corruption bound. The empirical results apply to these deployment assumptions
and the tested attack policies.

\section{Evaluation}
\label{sec:experiments}

We test how composed receipts change execution order, then examine whether
regional ordering can resist this attack while preserving normal precedence.

\subsection{Experimental Setup}
\label{subsec:attack-evaluation}

We replay Themis over latency measurements from 21 AWS
regions~\cite{park2025ambush} and 220 WonderProxy (WP)
locations~\cite{wonderproxy2020day}, following Ambush's trace-driven
methodology~\cite{park2025ambush}. Each full-graph configuration uses 400
workloads containing $T$, $F$, and 300 background transactions, with
directed delays estimated as RTT/2 in the baseline. Greedy training selects
corrupted locations from separate historical measurements and freezes
them before execution.

The committee-size experiments use $n\in\{9,21,101\}$, $f=(n-1)/4$,
and $\gamma=1$. Each corrupted replica observes the target and triggers
one co-located entry without added response delay, and the baseline quorum
includes all corrupted reporters. For mitigation, we use $n=21$ and five
co-located regions with one corrupted observer each.
Appendix~\ref{app:evaluation-details} gives the
report programs, regional thresholds, trace partitions, and replay statistics.

Because graph construction and finalization can give different pair
orders, we measure the \emph{strict-edge rate} before finalization and
the \emph{final attack rate} after Unspool. Their joint occurrence requires
both a strict $F\rightarrow T$ edge and final $F$-first execution. For
mitigation, we also report the \emph{direct-edge rate}, which includes
tie orientations before Unspool, and \emph{normal-order loss}, the fraction
of pairs with unanimous honest receipt order that are reversed in the
final order of attack-free workloads.

\subsection{Attack Effectiveness}
\label{subsec:attack-effectiveness}
\label{subsec:observer-composition}

\begin{figure}[!htbp]
  \centering
  \includegraphics[width=0.7\linewidth]{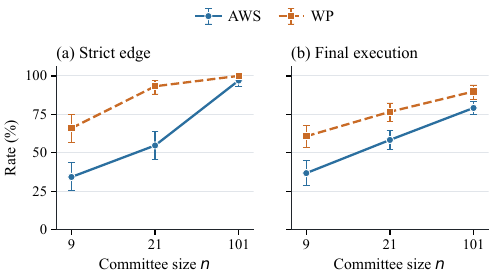}
  \caption{Attack effectiveness under plain dissemination: (a) strict-edge
  formation and (b) finalized overtaking. The corruption budget, active
  observers, and manipulated reports grow together: $f=2,5,25$.
  Error bars show 95\% confidence intervals.}
  \label{fig:attack-scaling}
\end{figure}

Themis finalizes the follower before its target at every tested committee
size, with WP providing more favorable response paths than AWS
(Figure~\ref{fig:attack-scaling}). Strict-edge formation and final execution
diverge because the latter includes tie orientations and Unspool can
reverse a pair's direct edge within an SCC. At $n=21$, each of the five
observers uses an entry at its own location, yielding final attack rates
of 58.3\% on AWS and 76.7\% on WP.

To isolate what submission paths add, we vary the total entry count $K$
while keeping all five observers and their complete corrupted reports
fixed. Each entry waits for observer notifications over measured links,
and the entry sets follow a nested sequence selected using historical
data. Increasing $K$ then raises both strict-edge formation and finalized
overtaking (Table~\ref{tab:client-entry-scaling}).
Appendix~\ref{app:client-entry-scaling} gives the single-observer results.

\begin{table}[!htbp]
  \centering
  \caption{Client-entry scaling with five fixed observers (\%).}
  \label{tab:client-entry-scaling}
  \footnotesize
  \setlength{\tabcolsep}{5pt}
  \begin{tabular}{@{}rrrrr@{}}
    \toprule
    & \multicolumn{2}{c}{AWS} & \multicolumn{2}{c}{WP} \\
    \cmidrule(lr){2-3}\cmidrule(l){4-5}
    $K$ & Strict edge & Final attack & Strict edge & Final attack \\
    \midrule
    1 & 13.50 & 36.88 & 36.25 & 50.00 \\
    5 & 39.75 & 50.76 & 76.25 & 68.19 \\
    20 & 62.75 & 61.50 & 87.00 & 73.96 \\
    \bottomrule
  \end{tabular}
\end{table}

\begin{figure}[!htbp]
  \centering
  \includegraphics[width=0.7\linewidth]{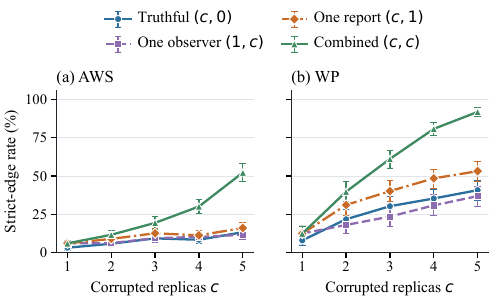}
  \caption{Capability ablation at $n=21$, $f=5$. Legend pairs count
  active observers and manipulated reports; $c$ is the actual corruption
  count. All configurations share the placement trained for $(c,c)$
  and matched replay inputs. Error bars show 95\% confidence intervals.}
  \label{fig:observer-composition}
\end{figure}

We then separate the effects of observation and report manipulation by
disabling each capability while retaining the trained placement
(Figure~\ref{fig:observer-composition}). Different entries
can still reach different honest replicas first when corrupted replicas
report truthfully, producing reverse edges from honest receipt support.
Manipulated reports add to that support, so the combined configuration
is strongest once multiple replicas are corrupted. Historical placement
also provides more support than random or regional-medoid placement,
as the matched comparison in
Appendix~\ref{app:evaluation-capability-placement} shows.

To compare combined paths with individual entries, we replay each
available entry separately with the same five manipulated reports,
target arrivals, and background workload, then take the idealized
per-workload maximum. The combined entries create a strict edge and
finalize $F$ first even when that best individual entry has insufficient
strict-edge support in 21.9\% of AWS workloads and 19.6\% of WP workloads.

\subsection{Attack Sensitivity}
\label{subsec:attack-sensitivity-study}

\begin{figure}[!htbp]
  \centering
  \includegraphics[width=0.7\linewidth]{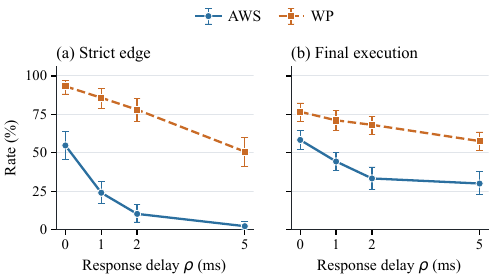}
  \caption{Response-delay sensitivity at $n=21$, $f=5$: (a) strict-edge
  formation and (b) finalized overtaking. Deployments, path delays,
  reporter identities, and background workloads are matched across offsets.
  Error bars show 95\% confidence intervals.}
  \label{fig:attack-sensitivity}
\end{figure}

\noindent\textbf{Response delay.}
The entry needs enough time to react before its copy loses the arrival
advantage. We vary the common additional processing delay $\delta$ from
Section~\ref{sec:attack-capacity}, labeled $\rho$ in
Figure~\ref{fig:attack-sensitivity}, to measure how quickly it consumes
the positive margins $M_r$ and removes honest supporters. AWS loses
strict-edge support more sharply than WP, while final attack rates
decline more slowly because tie orientations and SCC linearization can
still put $F$ first after strict-edge support is lost. A local response
benchmark and its timing breakdown appear in
Appendix~\ref{app:evaluation-local-processing}.

\noindent\textbf{Reporter selection.}
We next change whose arrivals supply the evidence, replacing corrupted
reporters with honest identities while keeping all five observers active.
Honest receipt support still leads to finalized overtaking after the
fabricated votes are removed (Table~\ref{tab:reporter-selection}). We also
let an honest leader collect the first complete reports to arrive, allowing
report propagation to select the quorum.

\begin{table}[!htbp]
  \centering
  \caption{Final attack rates under reporter selection (\%, $n=21$,
  $f=5$). All five observers remain active.}
  \label{tab:reporter-selection}
  \footnotesize
  \begin{tabular}{@{}lrr@{}}
    \toprule
    Reporter selection & AWS & WP \\
    \midrule
    All honest & 32.2 & 57.9 \\
    First arrivals at honest leader & 50.1 & 73.7 \\
    \bottomrule
  \end{tabular}
\end{table}

\noindent\textbf{Directional delays.}
The baseline estimates each directed delay as half of its measured RTT. To
examine that assumption, we use unequal directional shares and test both
the original placement and one retrained on the changed historical data
(Table~\ref{tab:directional-delay}). Since the target and follower traverse
different directed paths, these changes can create or remove positive
arrival margins $M_r$, which explains why the frozen AWS placement gains
ordering support under 20\% bias. Retraining then lets the attacker
select locations using the changed historical delays.

\begin{table}[!htbp]
  \centering
  \caption{Final attack rates under directional asymmetry (\%, $n=21$,
  $f=5$; 95\% CIs). Bias is the maximum deviation from RTT/2.}
  \label{tab:directional-delay}
  \footnotesize
  \setlength{\tabcolsep}{3pt}
  \begin{tabular}{@{}rlcc@{}}
    \toprule
    Max. bias & Placement & AWS & WP \\
    \midrule
    0 & Baseline & 58.3 [53.2, 63.3] & 76.7 [71.3, 82.0] \\
    20\% & Frozen & 67.2 [62.5, 72.0] & 76.6 [71.6, 81.5] \\
        & Retrained & 73.5 [68.8, 78.1] & 75.9 [69.9, 81.8] \\
    50\% & Frozen & 65.5 [60.7, 70.6] & 70.2 [64.9, 75.4] \\
         & Retrained & 69.4 [64.9, 73.9] & 73.6 [67.9, 79.1] \\
    \bottomrule
  \end{tabular}
\end{table}

We then use measured one-way-delay (OWD) distributions from ESnet's
public perfSONAR archive~\cite{esnetPerfsonar}. With $n=21$, $f=5$,
globally random honest-replica placement, and each observer's immediate
co-located submission upon receiving $T$, the attack forms a strict
$F\to T$ edge in 64.5\% of workloads and finalizes $F$ first in 69.3\%.
With five corrupted reports, each strict edge requires at least four of
the eleven selected honest reporters to receive $F$ first.
Appendix~\ref{app:evaluation-reporter-delay} describes the OWD sampling
and placement controls.

\noindent\textbf{Gossip.}
Replica forwarding changes when both observers and honest reporters
learn each transaction. On the original $n=21$ attack deployments, we
separate forwarding probability from batch waiting by increasing the
probability first, then forwarding each transaction immediately
(Table~\ref{tab:gossip-policies}). Removing the wait reduces overtaking
much more than increasing the probability with five-transaction batches.
All replicas follow the configured rule in the first four settings.
When corrupted replicas instead withhold $T$ while forwarding $F$, the
attack recovers much of its effectiveness on WP.

\begin{table}[!htbp]
  \centering
  \caption{Final attack rates under gossip (\%, $n=21$).}
  \label{tab:gossip-policies}
  \footnotesize
  \begin{tabular}{@{}lrr@{}}
    \toprule
    Forwarding policy & AWS & WP \\
    \midrule
    No gossip & 58.3 & 76.7 \\
    $p=0.1$, batches of five & 58.0 & 75.9 \\
    $p=1$, batches of five & 54.2 & 74.4 \\
    $p=1$, immediate & 23.6 & 20.8 \\
    $p=1$, immediate, selective & 33.3 & 63.9 \\
    \bottomrule
  \end{tabular}
\end{table}

Immediate forwarding lets $T$ and $F$ race over their flooding paths,
with $T$ beginning propagation before $F$ is released. This all-peer
flooding exchanges $O(n^2)$ transaction copies in a full peer mesh and
does not amortize message overhead across batches, so we use it as a
high-communication reference for dissemination effectiveness.
Appendix~\ref{app:evaluation-statistics} describes the event-driven replay.

\subsection{Mitigation Effectiveness}
\label{subsec:guard-evaluation}

The reference and adaptive replays use Algorithm~\ref{alg:regional-ordering}.
Every region has a complete base report, so fallback samples are nonempty.
Across 800 matched reference workloads, regional voting reduces overtaking
and normal-order loss relative to Themis on the same topology and gossip,
and to the uniform-random reference (Table~\ref{tab:regional-mitigation}).

\begin{table}[!htbp]
  \centering
  \caption{Mitigation effectiveness and normal-order loss (\%).}
  \label{tab:regional-mitigation}
  \footnotesize
  \setlength{\tabcolsep}{3pt}
  \begin{tabular}{@{}llrrr@{}}
    \toprule
    Method & Data & \shortstack{Direct\\$F\to T$}
      & \shortstack{Final\\attack rate}
      & \shortstack{Normal-order\\loss} \\
    \midrule
    \shortstack[l]{Uniform random\\total order} & Either & --- & 50 & 50 \\
    Themis (same gossip) & AWS & 30.25 & 42.65 & 2.09 \\
                         & WP & 9.00 & 29.90 & 5.09 \\
    Regional ordering & AWS & 11.07 & 20.05 & 0.52 \\
                      & WP & 0.75 & 16.60 & 1.54 \\
    \bottomrule
  \end{tabular}
\end{table}

On WP, most remaining successes occur even though regional voting
produces a $T\rightarrow F$ edge, because background transactions form
a return path that puts the pair in one SCC where Unspool executes $F$
first. Graph linearization also accounts for the observed normal-order
loss in attack-free workloads, since regional voting never reverses a
pair with unanimous honest receipt order in those runs.
Appendix~\ref{app:evaluation-regional-details} describes the attack-free
workloads.

\noindent\textbf{Adaptive testing.}
We hold the regional configuration fixed for all policies in
Table~\ref{tab:adapted-attacks}. The first two policies are preset. For the
third, we use historical data to select one policy per deployment and
evaluate it on held-out workloads.

\begin{table}[!htbp]
  \centering
  \caption{Final attack rates under adapted policies (\%).}
  \label{tab:adapted-attacks}
  \footnotesize
  \begin{tabular}{@{}lrr@{}}
    \toprule
    Attack policy & AWS & WP \\
    \midrule
    Early target relay & 17.7 & 16.6 \\
    Exchange and early relay & 17.5 & 16.2 \\
    Historical policy selection & 20.1 & 16.7 \\
    \bottomrule
  \end{tabular}
\end{table}

Relative to the fixed reference attack, historical selection changes the
final attack rate by $+0.02$ percentage points on AWS (paired 95\% CI
$[-0.79,0.91]$) and $+0.09$ points on WP ($[-0.19,0.53]$). Both intervals
include zero, so neither dataset shows a clear increase.

\FloatBarrier

\section{Related Work}
\label{sec:related-work}

\noindent\textbf{Receive-order fairness and its attacks.}
Order-fair consensus has been developed for permissioned and permissionless
settings~\cite{kelkar2020aequitas,kursawe2020wendy,cachin2022quickorderfairness,
zhang2020pompe,asayag2018helix,levari2019fairledger,kelkar2022permissionlessfairness}.
Later systems improve performance or change the source and aggregation of
ordering evidence~\cite{mu2024speedyfair,chen2024auncel,constantinescu2024dcn}.
Other fairness definitions compare differences in pairwise
support~\cite{cachin2022quickorderfairness}, restrict constraints to
data-dependent transactions~\cite{nagda2024rashnu}, or bound the extent of
an ordering reversal~\cite{kiayias2024boundedunfairness}.
FairDAG provides relative and absolute ordering variants atop multi-proposer
DAG consensus. DoD and Herring also integrate pairwise receipt aggregation
with DAG consensus~\cite{nagda2025dod,putnik2026herring,kang2025fairdag}.
AUTIG maintains incremental ordering graphs, while FEAT develops an
efficient transaction-ordering algorithm for $\gamma$-batch-order
fairness~\cite{ren2025autig,dinh2026feat}. Aequitas distinguishes
send-order fairness from guarantees based on actual receipts, noting the
additional timestamping and network assumptions needed for the former
~\cite{kelkar2020aequitas}. Themis explicitly studies network-level
front-running through triangle-inequality violations~\cite{kelkar2023themis}.
Our analysis combines different entries' favorable arrivals at different
recipients and follows their aggregate support through finalization.

Ambush is the closest attack study. It exploits temporary information
asymmetry to place a target and follower in the same Themis
component~\cite{park2025ambush}. The Condorcet attack shows that a malicious
client can deliberately induce cycles using legitimate
transactions~\cite{vafadar2023condorcet}. Lyra combines commit--reveal with
leaderless ordered consensus to resist transaction
reordering~\cite{zarbafian2023lyra}. These works study cycle formation or
reordering resistance, while our attack composes recipient-wise first arrivals
of copies sharing one transaction identifier into a direct reverse edge. We
also evaluate Ambush's delayed-gossip schedule in a matched replay at its
reported parameter point. Equal Opportunity studies systematic ordering bias
from infrastructure advantages and bounds it with secret
randomness~\cite{zhang2026equalopportunity}.

\noindent\textbf{Latency races and fair delivery.}
\emph{Buying Time} studies the economic competition between faster
delivery and bidding under transaction-ordering policies, and proposes
TimeBoost~\cite{mamageishvili2023buyingtime}. CloudEx uses clock
synchronization and buffering to compensate for delivery differences in
cloud-hosted exchanges~\cite{ghalayini2021cloudex}. DBO instead uses
logical delivery clocks to account for when participants received market
data~\cite{gupta2023dbo}. These systems make delivery conditions explicit
in the ordering rule, while our attack examines how multiple submission
entries affect receipt evidence in a Byzantine committee.

\noindent\textbf{Input causality.}
Ciampi et al.'s 2024 work formalizes sender-order fairness and input
causality in the universal-composition framework and gives a construction
using trusted enclaves~\cite{ciampi2024serialization}. Their 2025 work
combines its \emph{receiver-order fairness} notion with input causality using short-lived
transaction encryption, and also models transaction fees
~\cite{ciampi2025ucfairness}. Related encrypted
mempools hide transaction contents until ordering or batch
finalization~\cite{malkhi2022fino,zhang2023f3b,choudhuri2024bte,
choudhuri2025onetime,bormet2025beatmev}. Encryption prevents reactions
that require hidden transaction contents before ordering. Our main
model exposes the pending target only to corrupted replicas. PRECEDE takes an
incentive-based approach to causal ordering, using a power-weighted lottery to
make abstention an equilibrium response to a deterring bid~\cite{chen2026precede}.

\noindent\textbf{Constrained dissemination.}
L{\O} exchanges transaction commitments to make mempool behavior accountable,
while dissemination paths remain largely under miners'
control~\cite{nasrulin2023lo}. HERMES binds each message to a verifiably selected
random overlay and audits forwarding-path compliance~\cite{yahyaoui2025hermes}.
Each overlay uses $f+1$ entry points and redundant delivery paths.

\section{Conclusion}
We identify causal overtaking, in which a response executes before the
transaction that triggered its submission. We characterize how response
delays and multiple submission paths change the ordering support that
honest replicas provide, and apply the analysis to pairwise comparisons
across receive-order protocols. Under our network model, we prove that a
receipt-only rule cannot always preserve earlier-submission priority while
preventing causal overtaking. The proof constructs two histories with
identical signed receipt reports and opposite execution requirements. We
propose regional ordering as an empirical mitigation and evaluate it in
trace-driven simulations of Themis. In these simulations, overtaking occurs
even when all accepted reports come from honest replicas. With corruption
bounded within each region, regional ordering reduces overtaking in matched
replays and preserves most unanimous honest receipt orders in attack-free
workloads.

\section*{Ethical Considerations}
\phantomsection
\label{app:ethics}

Publishing these findings can help designers address causal overtaking before
deploying receive-order-fair protocols. We ran all attack experiments in
local simulations and have not identified an affected production deployment,
so we expect the immediate economic risk of publication to be limited.
Mitigation options include regional ordering, rules supported by trusted
submission timestamps or TEE-backed attestations, and encrypted mempools.
We plan to share our findings with the relevant protocol authors and seek
their feedback on deployment implications and possible mitigations.

\bibliographystyle{IEEEtran}
\bibliography{refs}

@inproceedings{werner2022sokdefi,
  title={Sok: Decentralized finance (defi)},
  author = {Werner, Sam and Perez, Daniel and Gudgeon, Lewis and Klages-Mundt, Ariah and Harz, Dominik and Knottenbelt, William},
  booktitle={Proceedings of the 4th ACM Conference on Advances in Financial Technologies},
  pages={30--46},
  year={2022}
}

@misc{esnetPerfsonar,
  author       = {{Energy Sciences Network}},
  title        = {Network Performance and Troubleshooting ({perfSONAR})},
  url          = {https://www.es.net/network-r-and-d/perfsonar/},
  note         = {Accessed: Nov. 17, 2026}
}

@inproceedings{heimbach2022sokreordering,
  title={Sok: Preventing transaction reordering manipulations in decentralized finance},
  author={Heimbach, Lioba and Wattenhofer, Roger},
  booktitle={Proceedings of the 4th ACM Conference on Advances in Financial Technologies},
  pages={47--60},
  year={2022}
}

@inproceedings{bano2019sokconsensus,
  title={SoK: Consensus in the age of blockchains},
  author={Bano, Shehar and Sonnino, Alberto and Al-Bassam, Mustafa and Azouvi, Sarah and McCorry, Patrick and Meiklejohn, Sarah and Danezis, George},
  booktitle={Proceedings of the 1st ACM Conference on Advances in Financial Technologies},
  pages={183--198},
  year={2019}
}

@inproceedings{gudgeon2020soklayer2,
  title={Sok: Layer-two blockchain protocols},
  author={Gudgeon, Lewis and Moreno-Sanchez, Pedro and Roos, Stefanie and McCorry, Patrick and Gervais, Arthur},
  booktitle={International Conference on Financial Cryptography and Data Security},
  pages={201--226},
  year={2020},
  organization={Springer}
}

@inproceedings{daian2020flashboys,
  title={Flash boys 2.0: Frontrunning in decentralized exchanges, miner extractable value, and consensus instability},
  author={Daian, Philip and Goldfeder, Steven and Kell, Tyler and Li, Yunqi and Zhao, Xueyuan and Bentov, Iddo and Breidenbach, Lorenz and Juels, Ari},
  booktitle={2020 IEEE symposium on security and privacy (SP)},
  pages={910--927},
  year={2020},
  organization={IEEE}
}

@inproceedings{torres2021frontrunnerjones,
  title={Frontrunner jones and the raiders of the dark forest: An empirical study of frontrunning on the ethereum blockchain},
  author = {Christof Ferreira Torres and Ramiro Camino and Radu State},
  booktitle={30th USENIX Security Symposium (USENIX Security 21)},
  pages={1343--1359},
  year={2021}
}

@misc{pornin2013deterministicecdsa,
  series =    {Request for Comments},
  number =    6979,
  howpublished =  {RFC 6979},
  publisher = {RFC Editor},
  doi =       {10.17487/RFC6979},
  url =       {https://www.rfc-editor.org/info/rfc6979/},
  author =    {T. Pornin},
  title =     {{Deterministic Usage of the Digital Signature Algorithm (DSA) and Elliptic Curve Digital Signature Algorithm (ECDSA)}},
  pagetotal = 79,
  year =      2013,
  month =     aug,
}

@inproceedings{kelkar2020aequitas,
  title={Order-fairness for byzantine consensus},
  author={Kelkar, Mahimna and Zhang, Fan and Goldfeder, Steven and Juels, Ari},
  booktitle={Annual International Cryptology Conference},
  pages={451--480},
  year={2020},
  organization={Springer}
}

@inproceedings{kelkar2023themis,
  title={Themis: Fast, strong order-fairness in byzantine consensus},
  author={Kelkar, Mahimna and Deb, Soubhik and Long, Sishan and Juels, Ari and Kannan, Sreeram},
  booktitle={Proceedings of the 2023 acm sigsac conference on computer and communications security},
  pages={475--489},
  year={2023}
}

@inproceedings{ciampi2024serialization,
  title={Universal composable transaction serialization with order fairness},
  author={Ciampi, Michele and Kiayias, Aggelos and Shen, Yu},
  booktitle={Annual International Cryptology Conference},
  pages={147--180},
  year={2024},
  organization={Springer}
}

@inproceedings{ciampi2025ucfairness,
  title={Universally composable transaction order fairness: Refined definitions and adaptive security},
  author={Ciampi, Michele and Kiayias, Aggelos and Shen, Yu},
  booktitle={International Conference on the Theory and Application of Cryptology and Information Security},
  pages={305--337},
  year={2025},
  organization={Springer}
}

@inproceedings{castro1999pbft,
  author = {Miguel Castro and Barbara Liskov},
title = {Practical Byzantine Fault Tolerance},
booktitle = {3rd Symposium on Operating Systems Design and Implementation (OSDI 99)},
year = {1999},
address = {New Orleans, LA},
url = {https://www.usenix.org/conference/osdi-99/practical-byzantine-fault-tolerance},
publisher = {USENIX Association},
month = feb
}

@inproceedings{yin2019hotstuff,
  title={HotStuff: BFT consensus with linearity and responsiveness},
  author={Yin, Maofan and Malkhi, Dahlia and Reiter, Michael K and Gueta, Guy Golan and Abraham, Ittai},
  booktitle={Proceedings of the 2019 ACM symposium on principles of distributed computing},
  pages={347--356},
  year={2019}
}

@incollection{lamport2019byzantine,
  title={The Byzantine generals problem},
  author={Lamport, Leslie and Shostak, Robert and Pease, Marshall},
  booktitle={Concurrency: the works of leslie lamport},
  pages={203--226},
  year={2019}
}

@article{schneider1990smr,
  title={Implementing fault-tolerant services using the state machine approach: A tutorial},
  author={Schneider, Fred B},
  journal={Acm Computing Surveys (CSUR)},
  volume={22},
  number={4},
  pages={299--319},
  year={1990},
  publisher={ACM, New York, NY, USA}
}

@article{dwork1988partialsynchrony,
  title={Consensus in the presence of partial synchrony},
  author={Dwork, Cynthia and Lynch, Nancy and Stockmeyer, Larry},
  journal={Journal of the ACM (JACM)},
  volume={35},
  number={2},
  pages={288--323},
  year={1988},
  publisher={ACM, New York, NY, USA}
}

@inproceedings{cachin2022quickorderfairness,
  title={Quick order fairness},
  author={Cachin, Christian and Mi{\'c}i{\'c}, Jovana and Steinhauer, Nathalie and Zanolini, Luca},
  booktitle={International Conference on Financial Cryptography and Data Security},
  pages={316--333},
  year={2022},
  organization={Springer}
}

@article{nagda2024rashnu,
  title={Rashnu: Data-dependent order-fairness},
  author={Nagda, Heena and Singhal, Shubhendra Pal and Amiri, Mohammad Javad and Loo, Boon Thau},
  journal={Proceedings of the VLDB Endowment},
  volume={17},
  number={9},
  year={2024},
  publisher={VLDB}
}

@inproceedings{mu2024speedyfair,
  author    = {Mu, Ke and Yin, Bo and Asheralieva, Alia and Wei, Xuetao},
  title={Separation is good: A faster order-fairness byzantine consensus},
  booktitle = {Proceedings of the Network and Distributed System Security Symposium},
  pages     = {1--17},
  year      = {2024},
  doi       = {10.14722/ndss.2024.24693},
  url       = {https://doi.org/10.14722/ndss.2024.24693}
}

@inproceedings{kursawe2020wendy,
  title={Wendy, the good little fairness widget: Achieving order fairness for blockchains},
  author={Kursawe, Klaus},
  booktitle={Proceedings of the 2nd ACM Conference on Advances in Financial Technologies},
  pages={25--36},
  year={2020}
}

@inproceedings{zhang2020pompe,
  title={Byzantine ordered consensus without byzantine oligarchy},
  author={Zhang, Yunhao and Setty, Srinath and Chen, Qi and Zhou, Lidong and Alvisi, Lorenzo},
  booktitle={14th USENIX Symposium on Operating Systems Design and Implementation (OSDI 20)},
  pages={633--649},
  year={2020}
}

@inproceedings{constantinescu2024dcn,
  author =	{Constantinescu, Andrei and Ghinea, Diana and Heimbach, Lioba and Wang, Zilin and Wattenhofer, Roger},
  title =	{{A Fair and Resilient Decentralized Clock Network for Transaction Ordering}},
  booktitle =	{27th International Conference on Principles of Distributed Systems (OPODIS 2023)},
  pages =	{8:1--8:20},
  series =	{Leibniz International Proceedings in Informatics (LIPIcs)},
  ISBN =	{978-3-95977-308-9},
  ISSN =	{1868-8969},
  year =	{2024},
  volume =	{286},
  editor =	{Bessani, Alysson and D\'{e}fago, Xavier and Nakamura, Junya and Wada, Koichi and Yamauchi, Yukiko},
  publisher =	{Schloss Dagstuhl -- Leibniz-Zentrum f{\"u}r Informatik},
  address =	{Dagstuhl, Germany},
  URL =		{https://drops.dagstuhl.de/entities/document/10.4230/LIPIcs.OPODIS.2023.8},
  URN =		{urn:nbn:de:0030-drops-194989},
  doi =		{10.4230/LIPIcs.OPODIS.2023.8}
}

@inproceedings{zarbafian2023lyra,
  title={Lyra: Fast and scalable resilience to reordering attacks in blockchains},
  author={Zarbafian, Pouriya and Gramoli, Vincent},
  booktitle={2023 IEEE International Parallel and Distributed Processing Symposium (IPDPS)},
  pages={929--939},
  year={2023},
  organization={IEEE}
}

@inproceedings{vafadar2023condorcet,
  author =	{Vafadar, Mohammad Amin and Khabbazian, Majid},
  title =	{{Condorcet Attack Against Fair Transaction Ordering}},
  booktitle =	{5th Conference on Advances in Financial Technologies (AFT 2023)},
  pages =	{15:1--15:21},
  series =	{Leibniz International Proceedings in Informatics (LIPIcs)},
  ISBN =	{978-3-95977-303-4},
  ISSN =	{1868-8969},
  year =	{2023},
  volume =	{282},
  editor =	{Bonneau, Joseph and Weinberg, S. Matthew},
  publisher =	{Schloss Dagstuhl -- Leibniz-Zentrum f{\"u}r Informatik},
  address =	{Dagstuhl, Germany},
  URL =		{https://drops.dagstuhl.de/entities/document/10.4230/LIPIcs.AFT.2023.15},
  URN =		{urn:nbn:de:0030-drops-192045},
  doi =		{10.4230/LIPIcs.AFT.2023.15}
}

@inproceedings{nasrulin2023lo,
  title={Lo: An accountable mempool for mev resistance},
  author={Nasrulin, Bulat and Ishmaev, Georgy and Decouchant, J{\'e}r{\'e}mie and Pouwelse, Johan},
  booktitle={Proceedings of the 24th international middleware conference},
  pages={98--110},
  year={2023}
}

@inproceedings{yahyaoui2025hermes,
  title={Mitigating front-running attacks through fair and resilient transaction dissemination},
  author={Yahyaoui, Wassim and Bruneau-Queyreix, Joachim and Decouchant, J{\'e}r{\'e}mie and V{\"o}lp, Marcus},
  booktitle={2025 55th Annual IEEE/IFIP International Conference on Dependable Systems and Networks (DSN)},
  pages={637--649},
  year={2025},
  organization={IEEE}
}

@inproceedings{park2025ambush,
  title={On frontrunning risks in batch-order fair systems for blockchains},
  author={Park, Eunchan and Yoon, Taeung and Nam, Hocheol and Maram, Deepak and Kang, Min Suk},
  booktitle={Proceedings of the 2025 ACM SIGSAC Conference on Computer and Communications Security},
  pages={918--932},
  year={2025}
}

@misc{wonderproxy2020day,
  author = {{WonderProxy}},
  title  = {A Day in the Life of the {Internet}},
  year   = {2020},
  month  = oct,
  url    = {https://wonderproxy.com/blog/a-day-in-the-life-of-the-internet/}
}

@article{kang2025fairdag,
  author = {Kang, Dakai and Chen, Junchao and Dinh, Tien Tuan Anh and Sadoghi, Mohammad},
  title = {FairDAG: Consensus Fairness over Multi-Proposer Causal Design},
  year = {2025},
  issue_date = {October 2025},
  publisher = {VLDB Endowment},
  volume = {19},
  number = {2},
  issn = {2150-8097},
  url = {https://doi.org/10.14778/3773749.3773763},
  doi = {10.14778/3773749.3773763},
  journal = {Proc. VLDB Endow.},
  month = oct,
  pages = {265--278},
  numpages = {14}
}

@inproceedings{dinh2026feat,
  title={FEAT: Fair and Efficient Adversarial Transaction Ordering},
  author={Dinh, Tien Tuan Anh and Kang, Dakai and Sadoghi, Mohammad},
  booktitle={ACM Symposium on Principles of Distributed Computing},
  pages={489--499},
  year={2026}
}

@article{chen2026precede,
  title={Time Is Money: Incentivized Causal Transaction Ordering},
  author={Chen, Hongyin and Zheng, Xu and Li, Jichen and Eyal, Ittay},
  journal={arXiv preprint arXiv:2607.11496},
  year={2026}
}

@inproceedings{levari2019fairledger,
  author =	{Lev-Ari, Kfir and Spiegelman, Alexander and Keidar, Idit and Malkhi, Dahlia},
  title =	{{FairLedger: A Fair Blockchain Protocol for Financial Institutions}},
  booktitle =	{23rd International Conference on Principles of Distributed Systems (OPODIS 2019)},
  pages =	{4:1--4:17},
  series =	{Leibniz International Proceedings in Informatics (LIPIcs)},
  ISBN =	{978-3-95977-133-7},
  ISSN =	{1868-8969},
  year =	{2020},
  volume =	{153},
  editor =	{Felber, Pascal and Friedman, Roy and Gilbert, Seth and Miller, Avery},
  publisher =	{Schloss Dagstuhl -- Leibniz-Zentrum f{\"u}r Informatik},
  address =	{Dagstuhl, Germany},
  URL =		{https://drops.dagstuhl.de/entities/document/10.4230/LIPIcs.OPODIS.2019.4},
  URN =		{urn:nbn:de:0030-drops-117904},
  doi =		{10.4230/LIPIcs.OPODIS.2019.4}
}

@inproceedings{zhang2026equalopportunity,
  title={Equal Opportunity: A Correctness Condition for Ordered Consensus},
  author={Zhang, Yunhao and Ni, Haobin and Basu, Soumya and Cohen, Shir and Yin, Maofan and Alvisi, Lorenzo and van Renesse, Robbert and Chen, Qi and Zhou, Lidong},
  booktitle={20th USENIX Symposium on Operating Systems Design and Implementation (OSDI 26)},
  pages={1283--1298},
  year={2026}
}

@inproceedings{kiayias2024boundedunfairness,
  title={Ordering transactions with bounded unfairness: Definitions, complexity and constructions},
  author={Kiayias, Aggelos and Leonardos, Nikos and Shen, Yu},
  booktitle={Annual International Conference on the Theory and Applications of Cryptographic Techniques},
  pages={34--63},
  year={2024},
  organization={Springer}
}

@inproceedings{kelkar2022permissionlessfairness,
  title={Order-fair consensus in the permissionless setting},
  author={Kelkar, Mahimna and Deb, Soubhik and Kannan, Sreeram},
  booktitle={Proceedings of the 9th ACM on ASIA Public-Key Cryptography Workshop},
  pages={3--14},
  year={2022}
}

@inproceedings{chen2024auncel,
  title={Auncel: Fair byzantine consensus protocol with high performance},
  author={Chen, Wuhui and Feng, Yikai and Zhang, Jianting and Cai, Zhongteng and Dai, Hong-Ning and Zheng, Zibin},
  booktitle={IEEE INFOCOM 2024-IEEE Conference on Computer Communications},
  pages={1251--1260},
  year={2024},
  organization={IEEE}
}

@inproceedings{asayag2018helix,
  title={A fair consensus protocol for transaction ordering},
  author={Asayag, Avi and Cohen, Gad and Grayevsky, Ido and Leshkowitz, Maya and Rottenstreich, Ori and Tamari, Ronen and Yakira, David},
  booktitle={2018 IEEE 26th International Conference on Network Protocols (ICNP)},
  pages={55--65},
  year={2018},
  organization={IEEE}
}

@inproceedings{mamageishvili2023buyingtime,
  author =	{Mamageishvili, Akaki and Kelkar, Mahimna and Schlegel, Jan Christoph and Felten, Edward W.},
  title =	{{Buying Time: Latency Racing vs. Bidding for Transaction Ordering}},
  booktitle =	{5th Conference on Advances in Financial Technologies (AFT 2023)},
  pages =	{23:1--23:22},
  series =	{Leibniz International Proceedings in Informatics (LIPIcs)},
  ISBN =	{978-3-95977-303-4},
  ISSN =	{1868-8969},
  year =	{2023},
  volume =	{282},
  editor =	{Bonneau, Joseph and Weinberg, S. Matthew},
  publisher =	{Schloss Dagstuhl -- Leibniz-Zentrum f{\"u}r Informatik},
  address =	{Dagstuhl, Germany},
  URL =		{https://drops.dagstuhl.de/entities/document/10.4230/LIPIcs.AFT.2023.23},
  URN =		{urn:nbn:de:0030-drops-192120},
  doi =		{10.4230/LIPIcs.AFT.2023.23}
}

@inproceedings{cachin2024qofimplementation,
  title={Quick order fairness: Implementation and evaluation},
  author={Cachin, Christian and Mi{\'c}i{\'c}, Jovana},
  booktitle={2024 IEEE International Conference on Blockchain and Cryptocurrency (ICBC)},
  pages={230--234},
  year={2024},
  organization={IEEE}
}

@inproceedings{cachin2001securebroadcast,
  title={Secure and efficient asynchronous broadcast protocols},
  author={Cachin, Christian and Kursawe, Klaus and Petzold, Frank and Shoup, Victor},
  booktitle={Annual International Cryptology Conference},
  pages={524--541},
  year={2001},
  organization={Springer}
}

@inproceedings{duan2017causalbroadcast,
  title={Secure causal atomic broadcast, revisited},
  author={Duan, Sisi and Reiter, Michael K and Zhang, Haibin},
  booktitle={2017 47th Annual IEEE/IFIP International Conference on Dependable Systems and Networks (DSN)},
  pages={61--72},
  year={2017},
  organization={IEEE}
}

@inproceedings{malkhi2022fino,
  author =	{Malkhi, Dahlia and Szalachowski, Pawel},
  title =	{{Maximal Extractable Value (MEV) Protection on a DAG}},
  booktitle =	{4th International Conference on Blockchain Economics, Security and Protocols (Tokenomics 2022)},
  pages =	{6:1--6:17},
  series =	{Open Access Series in Informatics (OASIcs)},
  ISBN =	{978-3-95977-274-7},
  ISSN =	{2190-6807},
  year =	{2023},
  volume =	{110},
  editor =	{Amoussou-Guenou, Yackolley and Kiayias, Aggelos and Verdier, Marianne},
  publisher =	{Schloss Dagstuhl -- Leibniz-Zentrum f{\"u}r Informatik},
  address =	{Dagstuhl, Germany},
  URL =		{https://drops.dagstuhl.de/entities/document/10.4230/OASIcs.Tokenomics.2022.6},
  URN =		{urn:nbn:de:0030-drops-184231},
  doi =		{10.4230/OASIcs.Tokenomics.2022.6}
}

@inproceedings{zhang2023f3b,
  author =	{Zhang, Haoqian and Merino, Louis-Henri and Qu, Ziyan and Bastankhah, Mahsa and Estrada-Gali\~{n}anes, Vero and Ford, Bryan},
  title =	{{F3B: A Low-Overhead Blockchain Architecture with Per-Transaction Front-Running Protection}},
  booktitle =	{5th Conference on Advances in Financial Technologies (AFT 2023)},
  pages =	{3:1--3:23},
  series =	{Leibniz International Proceedings in Informatics (LIPIcs)},
  ISBN =	{978-3-95977-303-4},
  ISSN =	{1868-8969},
  year =	{2023},
  volume =	{282},
  editor =	{Bonneau, Joseph and Weinberg, S. Matthew},
  publisher =	{Schloss Dagstuhl -- Leibniz-Zentrum f{\"u}r Informatik},
  address =	{Dagstuhl, Germany},
  URL =		{https://drops.dagstuhl.de/entities/document/10.4230/LIPIcs.AFT.2023.3},
  URN =		{urn:nbn:de:0030-drops-191921},
  doi =		{10.4230/LIPIcs.AFT.2023.3}
}

@inproceedings{choudhuri2024bte,
  title={Mempool privacy via batched threshold encryption: Attacks and defenses},
  author={Choudhuri, Arka Rai and Garg, Sanjam and Piet, Julien and Policharla, Guru-Vamsi},
  booktitle={33rd USENIX Security Symposium (USENIX Security 24)},
  pages={3513--3529},
  year={2024}
}

@inproceedings{choudhuri2025onetime,
  title={Practical mempool privacy via one-time setup batched threshold encryption},
  author={Choudhuri, Arka Rai and Garg, Sanjam and Policharla, Guru Vamsi and Wang, Mingyuan},
  booktitle={34th USENIX Security Symposium (USENIX Security 25)},
  pages={3477--3495},
  year={2025}
}

@inproceedings{bormet2025beatmev,
  author = {Jan Bormet and Sebastian Faust and Hussien Othman and Ziyan Qu},
title = {{BEAT-MEV}: Epochless Approach to Batched Threshold Encryption for {MEV} Prevention},
booktitle = {34th USENIX Security Symposium (USENIX Security 25)},
year = {2025},
isbn = {978-1-939133-52-6},
address = {Seattle, WA},
pages = {3457--3476},
url = {https://www.usenix.org/conference/usenixsecurity25/presentation/bormet},
publisher = {USENIX Association},
month = aug
}

@article{motepalli2023decentralizedsequencers,
  title={Sok: Decentralized sequencers for rollups},
  author={Motepalli, Shashank and Freitas, Luciano and Livshits, Benjamin},
  journal={arXiv preprint arXiv:2310.03616},
  year={2023}
}

@article{nagda2025dod,
  title={Dag of dags: Order-fairness made practical},
  author={Nagda, Heena and Sankhe, Sidharth and Sinha, Sakshi and Attarha, Keon and Amiri, Mohammad Javad and Loo, Boon Thau},
  journal={Proceedings of the ACM on Management of Data},
  volume={3},
  number={6},
  pages={1--27},
  year={2025},
  publisher={ACM, New York, NY, USA}
}

@article{putnik2026herring,
  title={Herring: Parallel Batch-Order-Fairness on DAG-based Blockchain Consensus},
  author={Putnik, Marko and Decouchant, J{\'e}r{\'e}mie},
  journal={arXiv preprint arXiv:2605.23648},
  year={2026}
}

@article{ren2025autig,
  title={Proof-Carrying Fair Ordering: Asymmetric Verification for BFT via Incremental Graphs},
  author={Ren, Pengkun and Dong, Hai and Sohrabi, Nasrin and Tari, Zahir and Zhang, Pengcheng},
  journal={arXiv preprint arXiv:2510.14186},
  year={2025}
}

@article{segalini2026tilikum,
  title={Tilikum: Transaction Fair Ordering on a DAG without Weak Edges},
  author={Segalini, Giulio and {\c{C}}olako{\u{g}}lu, Yigit and Putnik, Marko and Decouchant, J{\'e}r{\'e}mie},
  journal={arXiv preprint arXiv:2606.27250},
  year={2026}
}

@inproceedings{ramseyer2024streamingfairness,
  title={Brief announcement: Fair ordering via streaming social choice theory},
  author={Ramseyer, Geoffrey and Goel, Ashish},
  booktitle={Proceedings of the 43rd ACM Symposium on Principles of Distributed Computing},
  pages={279--282},
  year={2024}
}

@inproceedings{wang2007tiv,
  title={Towards network triangle inequality violation aware distributed systems},
  author = {Wang, Guohui and Zhang, Bo and Ng, T. S. Eugene},
  booktitle={Proceedings of the 7th ACM SIGCOMM conference on Internet measurement},
  pages={175--188},
  year={2007}
}

@inproceedings{ghalayini2021cloudex,
  title={CloudEx: a fair-access financial exchange in the cloud},
  author={Ghalayini, Ahmad and Geng, Jinkun and Sachidananda, Vighnesh and Sriram, Vinay and Geng, Yilong and Prabhakar, Balaji and Rosenblum, Mendel and Sivaraman, Anirudh},
  booktitle={Proceedings of the workshop on hot topics in operating systems},
  pages={96--103},
  year={2021}
}

@inproceedings{gupta2023dbo,
  title={DBO: Fairness for cloud-hosted financial exchanges},
  author={Gupta, Eashan and Goyal, Prateesh and Marinos, Ilias and Zhao, Chenxingyu and Mittal, Radhika and Chandra, Ranveer},
  booktitle={Proceedings of the ACM SIGCOMM 2023 Conference},
  pages={550--563},
  year={2023}
}

\appendix
\section{Evaluation Details}
\label{app:evaluation-details}

\subsection{Replay and Statistical Details}
\label{app:evaluation-statistics}

Full-graph replays use 20 deployments of 20 workloads per configuration
and average each workload over 20 public Unspool permutations shared
across paired runs.
For 95\% confidence intervals, we bootstrap deployments and workloads
hierarchically, computing paired differences before resampling.
Attack replays swap $T$ and $F$ in corrupted reports while preserving
background order. Complete reports bypass \textsc{FairUpdate}.
The capability ablation estimates strict-edge rates from 2,000 pair replays
and final attack rates from 400 full workloads.

Training and testing use disjoint measurements. AWS samples locations
globally, reusing the 21 profiles for $n=101$, while WP groups locations
by latency.

\textit{Gossip comparison.}
Each observer releases $F$ after actually receiving $T$ under the tested
forwarding rule. We count messages and model link delays, with no
bandwidth contention or congestion-dependent queueing.

\subsection{Reporter and Network-Delay Controls}
\label{app:evaluation-reporter-delay}

The honest leader takes the first 16 valid, complete reports to arrive.
The replay uses a common batch cutoff and measured reporter-to-leader
delays.

For each unordered location pair, a fixed $z\in[-1,1]$ gives directional
delays $\mathrm{RTT}(0.5\pm\beta z)$, using each direction's measured
RTT. The same field is used for training and evaluation at each bias,
and the intervals quantify workload variation within that field.

\textit{Measured one-way delays.}
We sample each message delay independently from ESnet's empirical OWAMP
distributions for 29 locations, using separate measurement windows for
training and testing. Clients can share locations with replicas.

\subsection{Capability and Placement}
\label{app:evaluation-capability-placement}

The capability ablation trains placement for $(c,c)$. Its one-observer
setting uses a location selected during training. In the matched
placement comparison, the trained, random, and regional-medoid rules give
strict-edge rates of
52.8\%, 16.5\%, and 3.5\% on AWS, and 93.2\%, 23.2\%, and 11.5\%
on WP, respectively.

\textit{Client-entry scaling.}
\label{app:client-entry-scaling}
The one-observer replay retains all five corrupted reports and uses the
same entry sequence as Table~\ref{tab:client-entry-scaling}.

\begin{table}[!htbp]
  \centering
  \caption{Client-entry scaling with one fixed observer (\%).}
  \label{tab:single-observer-entries}
  \footnotesize
  \setlength{\tabcolsep}{5pt}
  \begin{tabular}{@{}rrrrr@{}}
    \toprule
    & \multicolumn{2}{c}{AWS} & \multicolumn{2}{c}{WP} \\
    \cmidrule(lr){2-3}\cmidrule(l){4-5}
    $K$ & Strict edge & Final attack & Strict edge & Final attack \\
    \midrule
    1 & 10.25 & 34.53 & 32.25 & 47.94 \\
    5 & 15.25 & 35.15 & 51.00 & 57.38 \\
    20 & 15.25 & 35.23 & 52.50 & 59.09 \\
    \bottomrule
  \end{tabular}
\end{table}

\subsection{Local Processing Time}
\label{app:evaluation-local-processing}

On an idle ARM VM, we time 50,000 reactions from reading $T$ to writing
$F$ over loopback. The 99.9th percentile is 73.0\,$\mu$s with
construction and signing, and 31.7\,$\mu$s with a pre-signed follower.

\subsection{Regional-Ordering Replay Details}
\label{app:evaluation-regional-details}

The regions have sizes $5,4,4,4,4$ and use one measured location profile
each. A stable direction needs three witnesses. Otherwise, the region
samples up to three reports, with public priority resolving ties.
The sampling pool includes the complete base reports, which cover all
five regions in these replays. Three regional votes set the edge.
Each region reuses its two public priorities across transaction pairs.

The reference replay allows corrupted carriers to omit regional evidence
and corrupted reporters to claim a later target receipt. We weight
full-graph replays by the probabilities of the initial $T/F$ edge
directions. To measure normal-order loss, we remove $F$, restore truthful
target reports, and average reversal rates across workloads.

\section{Public Pending Reads}
\label{app:public-observation}

Consider a deployment where a public mempool or pending feed lets client
entries read transactions before ordering. Let
$d^{\mathsf{read}}(c_T,a)$ measure the elapsed time from $T$'s submission
until entry $a$ observes it through this interface, including the feed's
dissemination delay. After a local reaction delay $\rho_a$, entry $a$
submits $F$, giving the release offset
\[
  \alpha_a=d^{\mathsf{read}}(c_T,a)+\rho_a.
\]
Substituting this offset into the arrival and capacity expressions in
Section~\ref{sec:attack-capacity} accounts for direct observation by entries
in place of observer notifications.

With public observation, an instantiation can set $\mathcal B=\varnothing$
and $b=0$. The strict-edge condition becomes $v\geq q$: the accepted honest
reports alone must supply the required support for $F\rightarrow T$.
The trace evaluation in Section~\ref{sec:experiments} instantiates the
committee-observation model, where replicas supply these observations.

\section{Protocol Thresholds and Scope}
\label{app:protocol-thresholds}
\label{sec:protocol-case-studies}

\subsection{Native Support Thresholds}

We instantiate the threshold calculation in
Section~\ref{sec:attack-aggregation} for the eight constructions whose
stated comparisons satisfy all three properties in
Table~\ref{tab:protocol-coverage}. For a complete, admissible comparison,
write $s=b+v$ for forward support and $m-s$ for reverse support. The
resulting strict edge yields finalized overtaking under the execution
condition in that section.

For rules combining a support floor with a strict majority, abbreviate the
common expression as
$q_{\mathsf{floor}}(m,h)=\max\{\lceil h\rceil,\lfloor m/2\rfloor+1\}$.
Each instantiation specifies its native floor and fairness parameter.
The underlying definitions count either all replicas or honest replicas.

\noindent\textbf{Aequitas.}
The asynchronous construction at native $\gamma=1$ requires $n>4f$.
Its waiting graph adds $F\rightarrow T$ when at most $f$ logs in the union
of the two agreed log sets place $T$ at or before $F$
~\cite[full version, Sections~5.1 and~7]{kelkar2020aequitas}.
For complete strict orders this is $m-s\leq f$, or $s\geq m-f$.
With $m>2f$, forward support $s\geq m-f$ also excludes the reverse-edge
condition $s\leq f$.
Thus $q=m-f$. Unanimity at native $\gamma=1$ describes the fairness
antecedent over honest replicas, not the support needed among accepted
reports. The agreed logs must respect FIFO broadcast, and delivery
must satisfy the waiting-graph checks.
An observing corrupted replica may withhold $T$ until $F$ is ready, then
broadcast $F$ followed by $T$, provided this extends its accepted FIFO
prefix.

\noindent\textbf{Themis and SpeedyFair.}
Themis requires $s\geq\theta_\gamma$ and $s>m-s$ for a tie-independent
forward edge~\cite{kelkar2023themis}, with floor
$\theta_\gamma=n(1-\gamma)+\gamma f+1$ under the honest-replica fairness
parameterization of Section~\ref{sec:causal-goal}. Writing the native
all-replica parameter as $\gamma_{\mathsf{all}}$, the conversion is
$\gamma_{\mathsf{all}}n-f=\gamma(n-f)$. It converts the native floor
$n(1-\gamma_{\mathsf{all}})+f+1$ to $\theta_\gamma$ and the fault bound
to $n>2f(\gamma+1)/(2\gamma-1)$. At $\gamma=1$, this gives $n>4f$.
The strict-edge threshold is

\begin{equation}
\label{eq:themis-strict-edge}
  q_{\mathsf{strict}}(\mathcal S)
  :=q_{\mathsf{floor}}(m,\theta_\gamma).
\end{equation}

SpeedyFair collects $n-f$ signed local orders into a fragment and applies
\textsc{FairOrder}, combining Themis's proposal and update procedures
\cite[Sections~II-B and~IV]{mu2024speedyfair}. For the same complete input,
the pairwise threshold is unchanged. The local-order fragment is certified
before consensus commits the resulting order.

\noindent\textbf{Rashnu and DoD.}
Rashnu's initial construction applies the support floor and majority test
to data-dependent pairs~\cite[Algorithm~2, lines~10--14]{nagda2024rashnu}.
Write $h$ for its native floor $n(1-\gamma)+f+1$. Complete evidence gives
$q=q_{\mathsf{floor}}(m,h)$. Its later-update test uses a strict floor
comparison, so this reduction concerns the initial construction.

DoD uses the same two tests on $n-f$ local dependency graphs collected by
workers with the same ID across replicas
\cite[Algorithm~2, lines~11--16]{nagda2025dod}.
For a data-dependent pair in the same worker partition, with no existing
missing-edge history, its initial threshold is also
$q_{\mathsf{floor}}(m,h)$. Both transactions must survive admission and
enter the committed ordering graph.

\noindent\textbf{Herring.}
Herring constructs a graph from local-order snapshots associated with a
committed subdag~\cite[Algorithm~2]{putnik2026herring}.
Its initial pairwise rule selects the larger support when that support
reaches the native floor $h=n(1-\gamma)+f+1$.
For complete evidence with one comparison per replica, avoiding ties
therefore gives $q=q_{\mathsf{floor}}(m,h)$. Both transactions must remain
in the same graph after earlier assignments are filtered out and anchor
truncation is applied. This reduction addresses the initial comparison, before
any separate missing-edge resolution.

\noindent\textbf{FairDAG-RL.}
Section~6.2 of FairDAG aggregates committed local comparisons and orients
an unresolved pair according to their relative weights once its
quorum-based threshold is reached~\cite{kang2025fairdag}.
Consider a pair admitted to one graph with $m\geq n-f$ complete replica
comparisons from committed DAG data at its first resolution.
A strict majority $s>m-s$ implies
$s>m/2\geq(n-f)/2$, satisfying both the support test and the tie-independent
direction test. Hence $q=\lfloor m/2\rfloor+1$.

\noindent\textbf{Quick Order Fairness.}
QOF selects a cut specifying the consumed prefix of each sender's
FIFO-consistent broadcast log. Take honest broadcast order to follow
transaction input order. Let $w(F,T)$ and $w(T,F)$ denote the preferences
visible at this cut. Write $\zeta\geq0$ for QOF's fairness slack:
its guarantee applies when the honest preference margin exceeds
$2f+\zeta$~\cite{cachin2022quickorderfairness}.
QOF's forward-edge predicate is
\begin{equation}
\label{eq:qof-edge}
  \max\{w(F,T),\,n-f-w(T,F)\}>w(T,F)-f+\zeta.
\end{equation}
This predicate includes missing-evidence corrections
~\cite[Algorithm~2]{cachin2022quickorderfairness}.
If $m\geq n-f$ prefixes contain both transactions and all remaining
prefixes contain neither, then $w(F,T)=s$ and $w(T,F)=m-s$.
Since $n-f\leq m$, the maximum reduces to $s$, giving the forward-edge
condition $2s>m-f+\zeta$.
Applying the predicate in reverse and negating it gives
$2s\geq m+f-\zeta$.
The least integer satisfying both gives the strict-edge threshold:

\begin{equation}
\label{eq:qof-strict-threshold}
  \begin{aligned}
  q_{\mathsf{QOF}}=\max\biggl\{&
    \left\lfloor\frac{m-f+\zeta}{2}\right\rfloor+1,\\
    &\left\lceil\frac{m+f-\zeta}{2}\right\rceil
  \biggr\}.
  \end{aligned}
\end{equation}

Delivery of the resulting SCC additionally requires QOF's stability
test~\cite{cachin2024qofimplementation}. For prefixes containing only one
transaction, retain the missing-evidence terms in
Equation~\ref{eq:qof-edge}.

\subsection{Numerical Example}

For the admissible comparison cases above, take $n=21$, $f=5$, $m=16$,
and $\gamma=1$. Themis,
SpeedyFair, Rashnu, DoD, Herring, and FairDAG-RL require $q=9$ in the
complete comparison cases above. Asynchronous Aequitas at native
$\gamma=1$ and QOF at $\zeta=0$ instead require $q=11$.
With $b=5$, two entries winning at disjoint sets of three honest reporters
supply $v=6$, giving total support $s=b+v=11$. This reaches both
thresholds, so all eight specified comparisons create the strict edge.

\subsection{Other Aggregation Rules}
\label{app:aggregation-boundaries}

AUTIG adds $s$ and $m-s$ to the two cumulative weights and tests the
updated totals together with transaction states
~\cite[Section~4.3.2]{ren2025autig}. Tilikum-BOF requires the follower's
support to cross the threshold at a contiguous logical-counter prefix
where the reverse edge is still absent~\cite[Section~V-C]{segalini2026tilikum}.

Pomp\={e} orders commands by the median of $2f+1$ signed timestamp
responses~\cite{zhang2020pompe}. For $T$ timestamps $(5,20,10)$, follower
timestamps $(4,19,11)$ and $(4,9,11)$ both put $F$ first at the first two
responders, yet their medians lie on opposite sides of $T$'s median of 10.
An extension therefore needs the timestamp values and responder sets.

\end{document}